\documentclass[12pt,a4paper,american]{article}
\usepackage[utf8]{inputenc}
\usepackage[active]{srcltx}
\usepackage{amsthm}
\usepackage{amsmath}
\usepackage{graphicx}

\makeatletter

\numberwithin{equation}{section}
\numberwithin{figure}{section}
\newcommand{\lyxaddress}[1]{
\par {\raggedright #1
\vspace{1.4em}
\noindent\par}
}

\@ifundefined{date}{}{\date{}}

\usepackage{amsfonts}
\usepackage{amsthm}
\usepackage{mathrsfs}

\makeatletter
\newcommand{\manuallabel}[2]{\def\@currentlabel{#2}\label{#1}}
\makeatother
\newcommand{\sC}{\mathscr{C}}

\newtheorem*{proposition*}{Proposition}

\newtheorem{step}{Step}

\theoremstyle{remark}
\newtheorem*{remark*}{Remark}

\renewcommand{\Im}{\mathop\mathrm{Im}\nolimits}
\renewcommand{\Re}{\mathop\mathrm{Re}\nolimits}
\newcommand{\Dom}{\mathop\mathrm{Dom}\nolimits}

\newcommand{\dist}{\mathop\mathrm{dist}\nolimits}
\newcommand{\spec}{\mathop\mathrm{spec}\nolimits}

\makeatother

\usepackage{babel}
\begin{document}

\title{The heat kernel for two Aharonov-Bohm solenoids in a uniform magnetic
field}

\author{Pavel \v{S}\v{t}ov\'\i\v{c}ek}

\maketitle

\lyxaddress{Department of Mathematics, Faculty of Nuclear Science, Czech Technical
University in Prague, Trojanova 13, 120 00 Praha, Czech Republic}
\begin{abstract}
\noindent A non-relativistic quantum model is considered with a point
particle carrying a charge $e$ and moving on the plane pierced by
two infinitesimally thin Aharonov-Bohm solenoids and subjected to
a perpendicular uniform magnetic field of magnitude $B$. Relying
on a technique due to Schulman and Sunada which is applicable to Schrödinger
operators on multiply connected configuration manifolds a formula
is derived for the corresponding heat kernel. As an application of
the heat kernel formula, an approximate asymptotic expressions are
derived for the lowest eigenvalue lying above the first Landau level
and for the corresponding eigenfunction while assuming that $|eB|R^{2}/(\hbar c)$
is large where $R$ is the distance between the two solenoids. \vskip\baselineskip\noindent
\end{abstract}

\section{Introduction}

A non-relativistic quantum model is considered with a point particle
of mass $\mu$, carrying a charge $e$ and moving on the plane pierced
by two infinitesimally thin Aharonov-Bohm (AB) solenoids and subjected
to a perpendicular uniform magnetic field of magnitude $B$. The constants
$\mu$, $B$ are supposed positive, $e$ negative, and let
\[
\omega_{c}=\frac{|e|B}{\mu c}
\]
be the cyclotron frequency. Denote by $\Phi_{A}$, $\Phi_{B}$ be
the strengths of the two AB magnetic fluxes, and let
\begin{equation}
\alpha=-\frac{e\Phi_{A}}{2\pi\hbar c}\,,\ \beta=-\frac{e\Phi_{B}}{2\pi\hbar c}\,,\ \Lambda=(\alpha,\beta).\label{eq:PhiAB}
\end{equation}
This is a classical observation \cite{AharonovBohm} that the corresponding
magnetic Schrödinger operator $H_{\Lambda}$ is $1$-periodic both
in $\alpha$ and $\beta$ modulo unitary transformations. Thus we
can suppose, without loss of generality, that $\alpha,\beta\in(0,1)$.
Moreover, we assume that $\alpha\neq\beta$.

A thorough and detailed analysis of the spectral properties of $H_{\Lambda}$
due to T.~Mine can be found in \cite{mine}. The spectrum $\spec(H_{\Lambda})$
is known to be positive and pure point. It consists of infinitely
degenerate Landau levels $L_{n}=(n+1/2)\hbar\omega_{c}$, $n=0,1,2,\ldots$,
and of finitely degenerate isolated eigenvalues always located between
two neighboring Landau levels. In more detail, there are no spectral
points below $L_{0}$, and in each interval $(L_{n-1},L_{n})$, with
$n\in\mathbb{N}$, there are exactly 2$n$ eigenvalues (counted according
to their multiplicities) provided the distance $R$ between the solenoids
is sufficiently large. Moreover, $n$ of these eigenvalues are located
in a neighborhood of the value $(n+\alpha-1/2)\hbar\omega_{c}$ and
the other $n$ eigenvalues are located in a neighborhood of $(n+\beta-1/2)\hbar\omega_{c}$,
with the diameter of the neighborhoods shrinking with the rate at
least $\exp(-\delta R^{2})$ as $R$ increases where $\delta$ is
a positive constant.

The present paper pursues basically two goals. First, this is a construction
of the heat kernel for $H_{\Lambda}$ which is nothing but the integral
kernel of the semi-group of operators $\exp(-tH_{\Lambda}/\hbar)$,
$t>0$. Second, after having successfully accomplished the first task
one may wish to take advantage of the knowledge of the heat kernel
since it contains, in principle, all information about the spectral
properties of $H_{\Lambda}$. To extract this information one has
to rely on asymptotic analysis, as $t\to+\infty$, which may be technically
quite a difficult task. Assuming that $0<\alpha<\beta<1$, we focus
just on the simple eigenvalue located closely to $E_{1}:=(\alpha+1/2)\hbar\omega_{c}$.
This eigenvalue is called $E_{2}$ and it is the lowest eigenvalue
above the first Landau level $L_{0}$. We aim to construct an approximation
of the corresponding eigenfunction in the asymptotic domain $|e|BR^{2}/(\hbar c)\gg1$
and to determine the leading asymptotic term for the energy shift
$E_{2}-E_{1}$.

Concerning the first task we rely on a technique making it possible
to construct propagators or, similarly, heat kernels, on multiply
connected manifolds. This approach results in a formula, called the
Schulman-Sunada formula throughout the paper, which was derived by
Schulman in the framework of Feynman path integration \cite{schulman1,schulman2}
and, independently and mathematically rigorously, by Sunada \cite{sunada}.
In our model, the configuration manifold is a twice punctured plane.
A substantial step in this approach is a construction of the heat
kernel for a charged particle subjected to a uniform magnetic field
on the universal covering space of the configuration manifold. It
may be worthwhile emphasizing that, at this stage, no AB magnetic
fluxes are considered while working on the covering space; the AB
fluxes are incorporated into the model only after application of the
Schulman-Sunada formula.

A deeper understanding of the Schulman-Sunada formula may follow from
the fact that it can also be interpreted as an inverse procedure to
the Bloch decomposition of semi-bounded Schrödinger operators with
a discrete symmetry \cite{kocabovastovicek,kostakovastovicek}. The
technique has been used successfully, rather long time ago, in an
analysis of a similar model with two or more AB solenoids perpendicular
to the plane but without a uniform magnetic field in the background
\cite{pla89,jmp91}. Of course, the spectrum of the Hamiltonian in
this case is purely absolutely continuous and so the main focus in
such a model is naturally on the scattering problem \cite{pla91,pra93,dmj94}.
This quantum model has been approached also with some completely different
techniques like asymptotic methods for largely separated AB solenoids,
semiclassical analysis and a complex scaling method \cite{itotamura,tamura,alexandrovatamura,alexandrovatamura2}.
One may also mention more complex models comprising, apart of AB magnetic
fluxes, also additional potentials or magnetic fields \cite{mashkevichetal,mine},
or models with an arbitrary finite number of AB solenoids or even
with countably many solenoids arranged in a lattice \cite{dmj94,melgaardouhabazrozenblum,minenomura}.
On the other hand, the method stemming from the original ideas of
Schulman and Sunada turned out to be fruitful also in analysis of
other interesting models like Brownian random walk on the twice punctured
plane \cite{hannaythain,giraudetal}.

The present paper is organized as follows. To develop main ideas and
gain some experience we first treat, in Section~\ref{sec:One_AB_solenoid},
the case of a single solenoid embedded in a uniform magnetic field.
The heat kernel for the model with two AB solenoids in a uniform magnetic
field is derived in the form of an infinite series in Section~\ref{sec:The_two_solenoid_case}.
As an application of the heat kernel formula, we then focus on the
lowest simple eigenvalue of the magnetic Schrödinger operator which
lies above the first Landau level. First, in Section~\ref{sec:A_formula_eigenfunction},
an approximate formula is derived for a corresponding eigenfunction.
Then, in Section~\ref{sec:DeltaE}, this formula is used to obtain
an asymptotic approximation of the difference between the lowest eigenvalues
above the first Landau level for the cases of one and two AB solenoids
while assuming that $|e|BR^{2}/(\hbar c)$ is large. Some mathematical
technicalities are postponed to Appendices A, B and C.

The probability densities of the bound state in the model with one
or two solenoids, as mentioned above, are depicted on Figures \ref{fig:1solenoid}
and \ref{fig:2solenoid}, respectively. The former case is explicitly
solvable while in the latter case we use an approximate formula for
the eigenfunction in question which will be derived in Section~\ref{sec:A_formula_eigenfunction}.

\section{A single AB solenoid embedded in a uniform magnetic field \label{sec:One_AB_solenoid}}

\subsection{A uniform magnetic field on the universal covering space}

As a warm-up, we first study the model with one AB solenoid on the
background of a uniform magnetic field. This exercise should provide
us with some experience and clues how to proceed in the case of two
solenoids. The configuration manifold $M$ is the plane pierced at
one point, say the origin, $M=\mathbb{R}^{2}\setminus\{\pmb{0}\}\equiv(0,+\infty)\times S^{1}$
($S^{1}$ stands for the unit circle). For a electromagnetic vector
potential corresponding to a single AB solenoid we can choose
\[
\pmb{A}_{AB}=-\frac{\hbar c\alpha}{e}\,\nabla\theta
\]
where $\alpha$ is a parameter proportional to the strength of the
magnetic flux, see (\ref{eq:PhiAB}), and $(r,\theta)$ are polar
coordinates. Since the corresponding magnetic Schrödinger operators
for $\alpha$ and $\alpha+1$ are unitarily equivalent we shall suppose
here and everywhere in what follows that $\alpha\in(0,1)$. A uniform
magnetic field can be described by the vector potential (in symmetric
gauge) 
\begin{equation}
\pmb{A}_{B}=\frac{B}{2}\,\left(\!\begin{array}{c}
-x_{2}\\
x_{1}
\end{array}\!\right).\label{eq:avect_uni}
\end{equation}
The resulting magnetic Schrödinger operator in polar coordinates reads
\begin{equation}
H_{(B,\alpha)}=-\frac{\hbar^{2}}{2\mu}\!\left(\frac{\partial^{2}}{\partial r^{2}}+\frac{1}{r}\,\frac{\partial}{\partial r}+\frac{1}{r^{2}}\!\left(\frac{\partial}{\partial\theta}-i\,\frac{eBr^{2}}{2\hbar c}+i\alpha\right)^{\!2}\right)\!.\label{eq:H_B_alpha}
\end{equation}

The universal covering space of $M$ is $\tilde{M}=(0,+\infty)\times\mathbb{R}$.
Then $M=\tilde{M}/2\pi\mathbb{Z}$ (the structure group $\Gamma=2\pi\mathbb{Z}$
acts in the second factor of the Cartesian product). As the first
step, we consider a uniform magnetic field on $\tilde{M}$ only. An
AB magnetic flux will be incorporated into the model in the following
step, in Subsection~\ref{subsec:one_ABsolen}, after application
of the Schulman-Sunada formula. This means that at this stage we are
dealing with the Hamiltonian 
\[
\tilde{H}_{B}=-\frac{\hbar^{2}}{2\mu}\left(\frac{\partial^{2}}{\partial r^{2}}+\frac{1}{r}\,\frac{\partial}{\partial r}+\frac{1}{r^{2}}\!\left(\frac{\partial}{\partial\theta}-i\,\frac{eBr^{2}}{2\hbar c}\right)^{\!2}\right)
\]
which is supposed to act on $L^{2}(\tilde{M},r\mbox{d}r\mbox{d}\theta)$.
For a complete set of normalized generalized eigenfunctions of $\tilde{H}_{B}$
one can take $\{f_{n}(p;r,\theta);\ n\in\mathbb{Z}_{+},p\in\mathbb{R}\}$,
\begin{equation}
f_{n}(p;r,\theta)=C_{n}(p)\, r^{|p|}\, L_{n}^{|p|}\!\left(\frac{\mu\omega_{c}r^{2}}{2\hbar}\right)\exp\!\left(-\frac{\mu\omega_{c}r^{2}}{4\hbar}\right)e^{i\, p\theta},\label{eq:eigenfces}
\end{equation}
where $L_{n}^{\sigma}$ are Laguerre polynomials and 
\[
C_{n}(p)=\left(\frac{\mu\omega_{c}}{2\hbar}\right)^{\!(|p|+1)/2}\left(\frac{n!}{\pi\,\Gamma(n+|p|+1)}\right)^{\!1/2}\!.
\]
One has $\tilde{H}_{B}f_{n}(p)=\lambda_{n}(p)\, f_{n}(p)$ where
\begin{equation}
\lambda_{n}(p)=\left(\frac{p+|p|+1}{2}+n\right)\!\hbar\omega_{c}.\label{eq:eigenvals}
\end{equation}
The heat kernel on $\tilde{M}$ equals
\[
\tilde{p}_{t}(r,\theta;r_{0},\theta_{0})=\sum_{n\in\mathbb{Z}_{+}}\int_{\mathbb{R}}e^{-\lambda_{n}(p)t/\hbar}\, f_{n}(p;r,\theta)\,\overline{f_{n}(p;r_{0},\theta_{0})}\,\mbox{d}p.
\]

To proceed further, let us recall the Poisson kernel formula for Laguerre
polynomials \cite[Eq. (6.2.25)]{AndrewsAskeyRoy}: for $a,b,c,\sigma>0$,
\begin{eqnarray}
 &  & \sum_{n=0}^{\infty}e^{-cn}\,\frac{n!}{\Gamma(n+c+1)}\, L_{n}^{\sigma}(a)L_{n}^{\sigma}(b)\nonumber \\
 &  & =\,\frac{e^{\sigma c/2}}{(ab)^{\sigma/2}(1-e^{-c})}\,\exp\!\left(-\frac{(a+b)\, e^{-c}}{1-e^{-c}}\right)I_{\sigma}\!\left(\frac{2\sqrt{ab}\, e^{-c/2}}{1-e^{-c}}\right)\!.\label{eq:Poisson_kernel}
\end{eqnarray}
Applying (\ref{eq:Poisson_kernel}) one finds that
\begin{eqnarray}
\tilde{p}_{t}(r,\theta;r_{0},\theta_{0}) & = & \frac{\mu\omega_{c}}{4\pi\hbar\sinh(\omega_{c}t/2)}\,\exp\!\left(-\frac{\mu\omega_{c}\,(r^{2}+r_{0}^{\,2})}{4\hbar}\,\coth\!\left(\frac{\omega_{c}t}{2}\right)\right)\nonumber \\
 &  & \times\,\int_{\mathbb{R}}e^{-\left(\omega_{c}t/2+i\,(\theta-\theta_{0})\right)p}\, I_{|p|}\!\left(\frac{\mu\omega_{c}rr_{0}}{2\hbar\sinh(\omega_{c}t/2)}\right)\mbox{d}p.\label{eq:ptilde_1sol_Poisson}
\end{eqnarray}

Recall that the heat kernel for a charged particle on the plane in
a uniform magnetic field, if expressed in polar coordinates, reads
\cite[\S\  6.2.1.5]{GroscheSteiner}
\begin{eqnarray}
p_{t}(r,\theta;r_{0},\theta_{0}) & = & \frac{\mu\omega_{c}}{4\pi\hbar\sinh(\omega_{c}t/2)}\label{eq:heat_unim_polar}\\
 &  & \times\,\exp\!\left(-\frac{\mu\omega_{c}\,(r^{2}+r_{0}^{\,2})}{4\hbar}\,\coth\!\left(\frac{\omega_{c}t}{2}\right)+\frac{\mu\omega_{c}rr_{0}\cosh(\omega_{c}t/2-i(\theta-\theta_{0}))}{2\hbar\sinh(\omega_{c}t/2)}\right)\!.\nonumber 
\end{eqnarray}

This is the first opportunity in the present paper to demonstrate
how the Schulman-Sunada formula works. It claims that the heat kernels
$p_{t}$ and $\tilde{p}_{t}$ on $M$ and $\tilde{M}$, respectively,
are related by the equation
\begin{equation}
p_{t}(r,\theta;r_{0},\theta_{0})=\sum_{n\in\mathbb{Z}}\tilde{p}_{t}(r,\theta+2\pi n;r_{0},\theta_{0}).\label{eq:SuSch1}
\end{equation}
This is apparent, indeed, from the Poisson summation rule
\[
\sum_{n\in\mathbb{\mathbb{Z}}}\int_{\mathbb{R}}e^{2\pi inp}\,\varphi(p)\,\mbox{d}p=\sum_{n\in\mathbb{Z}}\varphi(n).
\]
In fact, (\ref{eq:SuSch1}) is then guaranteed by the identity \cite[Eq. 9.6.19]{AbramowitzStegun}
\[
\sum_{n\in\mathbb{Z}}e^{zn}\, I_{|n|}(x)=e^{x\cosh(z)},
\]
which is valid for $x>0$ and $z\in\mathbb{C}$.

Since $\tilde{M}$ looks locally like $M$ one may regard $p_{t}(r,\theta;r_{0},\theta_{0})$
as a $\Gamma$-invariant function on $\tilde{M}\times\tilde{M}$ which
is well defined provided $|\theta-\theta_{0}|<\pi$. Let us complete
$\tilde{M}$ with a point $A$ for which $r=0$ and $\theta$ is not
determined. The projection $\pi:\tilde{M}\to M$ extends so that $A$
projects onto the excluded point in $\mathbb{R}^{2}$ (the origin).
One has
\begin{equation}
p_{t}(A;r,\theta)=p_{t}(r,\theta;A)=\frac{\mu\omega_{c}}{4\pi\hbar\sinh(\omega_{c}t/2)}\,\exp\!\left(-\frac{\mu\omega_{c}\, r^{2}}{4\hbar}\,\coth\!\left(\frac{\omega_{c}t}{2}\right)\!\right)\!.\label{eq:heat_extremA}
\end{equation}

Being inspired by the approach applied in \cite{pla89} we wish to
express the heat kernel on $\tilde{M}$ in the form
\begin{eqnarray}
\tilde{p}_{t}(r,\theta;r_{0},\theta_{0}) & = & \vartheta(\pi-|\theta-\theta_{0}|)\, p_{t}(r,\theta;r_{0},\theta_{0})\label{eq:ptilde_1sol_gen}\\
 &  & +\,\int_{0}^{t}p_{s}(r,\theta;A)V(r,\theta,s;r_{0},\theta_{0},t-s)\, p_{t-s}(A;r_{0},\theta_{0})\,\mbox{d}s.\nonumber 
\end{eqnarray}
Here $\vartheta$ stands for the Heaviside step function ($\vartheta(x)$
equals $1$ for $x>0$ and $0$ otherwise). Concerning the angles,
the expression $V(r,\theta,s;r_{0},\theta_{0},t-s)$ should depend
only on their difference $\theta-\theta_{0}$. To obtain the sought
expression one can use the formula ($x>0$) \cite[Eq. 9.6.20]{AbramowitzStegun}
\[
I_{\nu}(x)=\frac{1}{\pi}\,\int_{0}^{\pi}e^{x\cos(\phi)}\,\cos(\nu\phi)\,\mbox{d}\phi-\frac{\sin(\nu\pi)}{\pi}\,\int_{0}^{\infty}e^{-x\cosh(s)-\nu s}\,\mbox{d}s.
\]
We find that, for any $z\in\mathbb{C}$,
\begin{eqnarray}
\int_{-\infty}^{\infty}e^{zp}\, I_{|p|}(x)\,\mbox{d}p & = & e^{x\cosh(z)}\,\vartheta(\pi-|\Im z|)\label{eq:int_IBessel}\\
 &  & +\,\frac{1}{2\pi i}\int_{-\infty}^{\infty}e^{-x\cosh(u)}\left(\frac{1}{z+u+i\pi}-\frac{1}{z+u-i\pi}\right)\!\mbox{d}u.\nonumber 
\end{eqnarray}

The identity (\ref{eq:int_IBessel}) is to be applied for $z=-\omega_{c}t/2+i\,(\theta-\theta_{0})$,
$x=\mu\omega_{c}rr_{0}/(2\hbar\sinh(\omega_{c}t/2))$. In combination
with (\ref{eq:ptilde_1sol_Poisson}) we obtain
\begin{eqnarray}
 &  & \hskip-2em\tilde{p}_{t}(r,\theta;r_{0},\theta_{0})\,=\,\vartheta(\pi-|\theta-\theta_{0}|)\, p_{t}(r,\theta;r_{0},\theta_{0})\nonumber \\
\noalign{\smallskip} &  & \qquad+\,\frac{\mu\omega_{c}}{4\pi\hbar\sinh(\omega_{c}t/2)}\,\exp\!\left(-\frac{\mu\omega_{c}\,(r^{2}+r_{0}^{\,2})}{4\hbar}\,\coth(\omega_{c}t/2)\right)\label{eq:ptilde_1sol_I_int}\\
\noalign{\smallskip} &  & \qquad\quad\times\,\frac{1}{2\pi i}\int_{-\infty}^{\infty}\exp\!\left(-\frac{\mu\omega_{c}rr_{0}}{2\hbar\sinh(\omega_{c}t)}\,\cosh(u)\right)\nonumber \\
 &  & \qquad\qquad\qquad\qquad\times\left(\frac{1}{u-\omega_{c}t/2+i\,(\theta-\theta_{0}+\pi)}-\frac{1}{u-\omega_{c}t/2+i\,(\theta-\theta_{0}-\pi)}\right)\mbox{d}u.\nonumber 
\end{eqnarray}
To bring the formula to the desired structure we seek a substitution
$u=u(s)$ such that
\begin{eqnarray*}
 &  & \frac{\mu\omega_{c}\,(r^{2}+r_{0}^{\,2})}{4\hbar}\,\coth\!\left(\frac{\omega_{c}t}{2}\right)+\frac{\mu\omega_{c}rr_{0}}{2\hbar\sinh(Bt)}\,\cosh(u)\\
 &  & =\,\frac{\mu\omega_{c}r^{2}}{4\hbar}\,\coth\!\left(\frac{\omega_{c}s}{2}\right)+\frac{\mu\omega_{c}r_{0}^{\,2}}{4\hbar}\,\coth\!\left(\frac{\omega_{c}(t-s)}{2}\right).
\end{eqnarray*}
Simple manipulations lead to the solution
\[
u(s)=\ln\!\left(\frac{r\sinh\!\big(\omega_{c}(t-s)/2\big)}{r_{0}\sinh(\omega_{c}s/2)}\right)\!.
\]
Then the second term on the RHS of (\ref{eq:ptilde_1sol_I_int}) equals
\begin{eqnarray*}
\noalign{\medskip} &  & \hskip-2em\frac{\mu}{i\hbar}\left(\frac{\omega_{c}}{4\pi}\right)^{\!2}\int_{0}^{t}\frac{\exp\!\left(-\mu\omega_{c}r^{2}\coth(\omega_{c}s/2)/(4\hbar)\right)}{\sinh(\omega_{c}s/2)}\,\frac{\exp\!\left(-\mu\omega_{c}r_{0}^{\,2}\coth\big(\omega_{c}\,(t-s)/2\big)/(4\hbar)\right)}{\sinh\big(\omega_{c}\,(t-s)/2\big)}\\
\noalign{\medskip} &  & \qquad\qquad\times\!\left(\frac{1}{u(s)-\omega_{c}t/2+i\,(\theta-\theta_{0}+\pi)}-\frac{1}{u(s)-\omega_{c}t/2+i\,(\theta-\theta_{0}-\pi)}\right)\!\mbox{d}s.
\end{eqnarray*}

We conclude that (\ref{eq:ptilde_1sol_gen}) holds, indeed, provided
we set
\begin{eqnarray}
\hskip-1emV(r,\theta,s;r_{0},\theta_{0},s_{0}) & = & \frac{\hbar}{i\mu}\,\Bigg(\!\left(U(r,s;r_{0},s_{0})-\frac{\omega_{c}\,(s+s_{0})}{2}+i\,(\theta-\theta_{0}+\pi)\right)^{\!-1}\nonumber \\
\hskip-1em &  & \qquad-\left(U(r,s;r_{0},s_{0})-\frac{\omega_{c}\,(s+s_{0})}{2}+i\,(\theta-\theta_{0}-\pi)\right)^{\!-1}\Bigg)\nonumber \\
\label{eq:V_1sol}
\end{eqnarray}
where
\begin{equation}
U(r,s;r_{0},s_{0})=\ln\!\left(\frac{r\sinh(\omega_{c}s_{0}/2)}{r_{0}\sinh(\omega_{c}s/2)}\right)\!.\label{eq:V_U_aux}
\end{equation}

\subsection{An application of the Schulman-Sunada formula \label{subsec:one_ABsolen}}

If an AB flux is switched on the Schulman-Sunada formula (\ref{eq:SuSch1})
should be modified. In that case it claims that the heat kernel of
the magnetic Schrödinger operator corresponding to one AB solenoid
in a uniform magnetic field on the plane equals
\begin{equation}
p_{t}^{(\alpha)}(r,\theta;r_{0},\theta_{0})=\sum_{n\in\mathbb{Z}}e^{2\pi i\alpha n}\,\tilde{p}_{t}(r,\theta-2\pi n;r_{0},\theta_{0}).\label{eq:SuSch2}
\end{equation}

$p_{t}^{(\alpha)}(r,\theta;r_{0},\theta_{0})$ can be interpreted
as an equivariant function on $\tilde{M}$ or, alternatively, as a
multivalued function on $M$. To obtain an unambiguous expression
one has to choose a cut on $M$. We restrict ourselves to values $\theta\in(-\pi,\pi)$,
$\theta_{0}=0$. Note that, under these restrictions, $\vartheta(\pi-|\theta-2\pi n-\theta_{0}|)\neq0$
if and only if $n=0$. We shall need the readily verifiable identity
\begin{equation}
\sum_{n=-\infty}^{\infty}\frac{e^{2\pi i\alpha n}}{w-2\pi n}=\frac{i\, e^{i\alpha w}}{e^{iw}-1}\label{eq:sum_aux1}
\end{equation}
which holds for $\alpha\in(0,1)$ and $w\in\mathbb{C}\setminus2\pi\mathbb{Z}$.
Plugging (\ref{eq:ptilde_1sol_I_int}) into (\ref{eq:SuSch2}) and
using (\ref{eq:sum_aux1}) we obtain
\begin{eqnarray}
 &  & \hskip-1.5emp_{t}^{(\alpha)}(r,\theta;r_{0},0)=\frac{\mu\omega_{c}}{4\pi\hbar\sinh(\omega_{c}t/2)}\nonumber \\
 &  & \qquad\qquad\qquad\qquad\times\exp\!\left(-\frac{\mu\omega_{c}\,(r^{2}+r_{0}^{\,2})}{4\hbar}\,\coth\!\left(\frac{\omega_{c}t}{2}\right)\!+\frac{\mu\omega_{c}rr_{0}\cosh(\omega_{c}t/2-i\theta)}{2\hbar\sinh(\omega_{c}t/2)}\right)\nonumber \\
\noalign{\smallskip} &  & \qquad\qquad\qquad-\,\frac{\sin(\pi\alpha)\,\mu\omega_{c}}{4\pi^{2}\hbar\sinh(\omega_{c}t/2)}\,\exp\!\left(-\frac{\mu\omega_{c}\,(r^{2}+r_{0}^{\,2})}{4\hbar}\,\coth(\omega_{c}t/2)\right)\label{eq:palpha_1sol_intu}\\
 &  & \qquad\qquad\qquad\qquad\times\int_{\mathbb{R}}\exp\!\left(-\frac{\mu\omega_{c}rr_{0}\cosh(u+\omega_{c}t/2)}{2\hbar\sinh(\omega_{c}t/2)}\right)\!\frac{e^{\alpha(u+i\theta)}}{1+e^{u+i\theta}}\,\mbox{d}u.\nonumber 
\end{eqnarray}

A common approach how to derive a formula for $p_{t}^{(\alpha)}(r,\theta;r_{0},0)$
is based on the eigenfunction expansion. Referring to (\ref{eq:eigenfces}),
(\ref{eq:eigenvals}), the functions
\begin{equation}
f_{n,m}(r,\theta):=f_{n}(m+\alpha;r,\theta),\ \ \text{with}\ n\in\mathbb{Z}_{+},m\in\mathbb{Z},\label{eq:eigenfce_H_Ba}
\end{equation}
form an orthonormal basis in $L^{2}(M,r\mbox{d}r\mbox{d}\theta)$.
At the same time, the functions
\[
f'_{n,m}(r,\theta)=e^{-i\alpha\theta}f{}_{n,m}(r,\theta)
\]
are eigenfunctions of $H_{(B,\alpha)}$ (see (\ref{eq:H_B_alpha})),
$H_{(B,\alpha)}f'_{n,m}=\lambda_{n,m}f'_{n,m}$ where
\begin{equation}
\lambda_{n,m}:=\lambda_{n}(m+\alpha)=\!\left(\frac{m+\alpha+|m+\alpha|+1}{2}+n\right)\!\hbar\omega_{c}.\label{eq:lbd_nm}
\end{equation}
Then
\begin{equation}
p_{t}^{(\alpha)}(r,\theta;r_{0},0)=\sum_{n\in\mathbb{Z}_{+}}\sum_{m\in\mathbb{Z}}\, e^{-t\lambda_{n,m}/\hbar}f_{n,m}(r,\theta)\,\overline{f_{n,m}(r_{0},0)}\,.\label{eq:palpha_1sol_egexpand}
\end{equation}

Conversely, (\ref{eq:palpha_1sol_egexpand}) suggests that $p_{t}^{(\alpha)}$
contains all information about the spectral properties of $H_{(B,\alpha)}$
which can be extracted, in principle, with the aid of asymptotic analysis
of the heat kernel as $t\to+\infty$. As an example, let us check
just the first two leading asymptotic terms of (\ref{eq:palpha_1sol_intu}).
Referring to (\ref{eq:asympt_single}), it is straightforward to derive
the asymptotic expansion
\begin{eqnarray}
 &  & \hskip-2emp_{t}^{(\alpha)}(r,\theta;r_{0},0)\,=\Bigg(\frac{\mu\omega_{c}}{2\pi\hbar}\,\exp\!\left(-\frac{\mu\omega_{c}\,(r^{2}+r_{0}^{\,2})}{4\hbar}+\frac{\mu\omega_{c}rr_{0}}{2\hbar}\, e^{-i\theta}\right)\nonumber \\
\noalign{\smallskip} &  & -\,\frac{\sin(\pi\alpha)\mu\omega_{c}}{2\pi^{2}\hbar}\,\exp\!\left(-\frac{\mu\omega_{c}\,(r^{2}+r_{0}^{\,2})}{4\hbar}\right)\int_{\mathbb{R}}\exp\!\left(-\frac{\mu\omega_{c}rr_{0}}{2\hbar}\, e^{u}\right)\frac{e^{\alpha(u+i\theta)}}{1+e^{u+i\theta}}\,\mbox{d}u\Bigg)e^{-\omega_{c}t/2}\nonumber \\
 &  & +\,\frac{\mu\omega_{c}}{2\pi\Gamma(1+\alpha)\hbar}\,\exp\!\left(-\frac{\mu\omega_{c}\,(r^{2}+r_{0}^{\,2})}{4\hbar}\right)\left(\frac{\mu\omega_{c}rr_{0}}{2\hbar}\right)^{\!\alpha}e^{i\alpha\theta}\, e^{-(\alpha+1/2)\omega_{c}t}+O(e^{-3\omega_{c}t/2}).\nonumber \\
\noalign{\vskip-10pt}\label{eq:palpha_1sol_asympt}
\end{eqnarray}

Comparing (\ref{eq:palpha_1sol_egexpand}) with (\ref{eq:palpha_1sol_asympt}),
the asymptotic expansion covers the terms in the former formula with
$n=0$ and $m\in-\mathbb{Z}_{+}$. For $n=0$ and $m=0$ one reveals
the expression for the eigenfunction $f_{0,0}(r,\theta)$ corresponding
to the simple eigenvalue $\lambda_{0,0}=(\alpha+1/2)\hbar\omega_{c}$,
see (\ref{eq:eigenfce_H_Ba}), (\ref{eq:lbd_nm}).

For $n=0$ and $m<0$, $\lambda_{0,m}=\hbar\omega_{c}/2$ is the infinitely
degenerate lowest Landau level. Hence the expression standing at $e^{-\omega_{c}t/2}$
on the RHS of (\ref{eq:palpha_1sol_asympt}) should be the kernel
of the orthogonal projection onto the corresponding eigenspace. This
is to say that
\begin{eqnarray}
 &  & \hskip-2em\sum_{m\in\mathbb{Z},\, m<0}f_{0,m}(r,\theta)\,\overline{f_{0.m}(r_{0},0)}\,=\,\frac{\mu\omega_{c}}{2\pi\hbar}\,\exp\!\left(-\frac{\mu\omega_{c}\,(r^{2}+r_{0}^{\,2})}{4\hbar}+\frac{\mu\omega_{c}rr_{0}}{2\hbar}\, e^{-i\theta}\right)\nonumber \\
 &  & \qquad\qquad-\,\frac{\sin(\pi\alpha)\mu\omega_{c}}{2\pi^{2}\hbar}\,\exp\!\left(-\frac{\mu\omega_{c}\,(r^{2}+r_{0}^{\,2})}{4\hbar}\right)\int_{\mathbb{R}}\exp\!\left(-\frac{\mu\omega_{c}rr_{0}}{2\hbar}\, e^{u}\right)\frac{e^{\alpha(u+i\theta)}}{1+e^{u+i\theta}}\,\mbox{d}u.\nonumber \\
\noalign{\vskip-5pt}\label{eq:project_n0_eq1}
\end{eqnarray}
Observing that $L_{0}^{\sigma}(x)=1$, (\ref{eq:project_n0_eq1})
amounts to
\begin{equation}
\exp(\rho e^{i\phi})\, e^{i\alpha\phi}-\frac{\sin(\pi\alpha)}{\pi}\int_{\mathbb{R}}\exp(-\rho e^{u})\frac{e^{\alpha u}}{1+e^{u-i\phi}}\,\mbox{d}u=\sum_{m=1}^{\infty}\frac{\rho^{m-\alpha}}{\Gamma(m+1-\alpha)}\, e^{im\phi}\label{eq:four_ser_aux1}
\end{equation}
for $\rho>0$ and $\phi\in(-\pi,\pi)$ (and $\alpha\in(0,1)$). But
(\ref{eq:four_ser_aux1}) follows from (\ref{eq:Gamma_incompl3}).

\section{The two-solenoid case \label{sec:The_two_solenoid_case}}

\subsection{The heat kernel on the universal covering space}

The configuration space for the Aharonov-Bohm effect with two vortices
is the plane with two excluded points, $M=\mathbb{R}^{2}\setminus\{\pmb{a},\pmb{b}\}$.
$M$ is a flat Riemannian manifold and the same is true for the universal
covering space $\tilde{M}$. By the very construction, $\tilde{M}$
is connected and simply connected, and one can identify $M=\tilde{M}/\Gamma$
where $\Gamma$ is the fundamental group of $M$. Let $\pi:\tilde{M}\to M$
be the projection.

The fundamental group $\Gamma$ in this case is known to be the free
group with two generators. For the first generator, called $g_{a}$,
one can choose the homotopy class of a simple positively oriented
loop winding once around the point $\pmb{a}$ and leaving the point
$\pmb{b}$ in the exterior. Analogously one can choose the second
generator, $g_{b}$, by interchanging the role of $\pmb{a}$ and $\pmb{b}$.

It is convenient to complete the manifold $\tilde{M}$ by a countable
set of points $\mathcal{A}\cup\mathcal{B}$ lying on the border of
$\tilde{M}$ and projecting onto the excluded points, $\pi(\mathcal{A})=\{\pmb{a}\}$
and $\pi(\mathcal{B})=\{\pmb{b}\}$.

The geometry of $\tilde{M}$ is locally the same as that of the plane
but globally it exhibits some substantially distinct features. In
particular, not always a couple of points from $\tilde{M}$ can be
connected by a segment (a geodesic curve). For $\pmb{x},\pmb{y}\in\tilde{M}\cup\mathcal{A}\cup\mathcal{B}$
set $\chi(\pmb{x},\pmb{y})=1$ if the points $\pmb{x}$, $\pmb{y}$
can be connected by a segment, and $\chi(\pmb{x},\pmb{y})=0$ otherwise.

Our first goal is to construct the heat kernel on $\tilde{M}$ for
a charged particle subjected to a uniform magnetic field. A starting
point is again formula (\ref{eq:heat_unim_polar}) for the heat kernel
on the plane which can be rewritten in Cartesian coordinates as
\begin{equation}
p_{t}(\pmb{x},\pmb{x}_{0})=\frac{\mu\omega_{c}}{4\pi\hbar\sinh(Bt)}\,\exp\!\left(-\frac{\mu\omega_{c}\,|\pmb{x}-\pmb{x}_{0}|^{2}}{4\hbar}\,\coth\!\left(\frac{\omega_{c}t}{2}\right)+i\,\frac{\mu\omega_{c}}{2\hbar}\,\pmb{x}\wedge\pmb{x}_{0}\right)\!.\label{eq:heat_ker_Cartes}
\end{equation}
Here we have used the notation $\pmb{x}\wedge\pmb{y}:=x_{1}y_{2}-x_{2}y_{1}$.
Note that $\pmb{x}\wedge\pmb{x}_{0}=-rr_{0}\sin(\theta-\theta_{0})$.

Our choice of the gauge for the electromagnetic vector potential depends
on the choice of the origin of coordinates. Moving the origin of coordinates
from $\pmb{0}$ to a point $\pmb{y}$ one has the transformation rule
\begin{equation}
p_{t}^{\text{transformed}}(\pmb{x};\pmb{x}_{0})=e^{i\mu\omega_{c}/(2\hbar)\,\pmb{y}\wedge\pmb{x}}\, p_{t}(\pmb{x};\pmb{x}_{0})\, e^{-i\mu\omega_{c}/(2\hbar)\,\pmb{y}\wedge\pmb{x}_{0}}.\label{eq:pt_transform}
\end{equation}
For the corresponding Hamiltonian, now written in Cartesian coordinates
as
\begin{equation}
H_{B}=-\frac{\hbar^{2}}{2\mu}\left(\frac{\partial}{\partial x_{1}}+\frac{ieB}{2\hbar c}\, x_{2}\right)^{\!2}-\left(\frac{\partial}{\partial x_{2}}-\frac{ieB}{2\hbar c}\, x_{1}\right)^{\!2},\label{eq:HB_Cartesian}
\end{equation}
this means the unitary transformation 
\[
H_{B}^{\text{transformed}}=e^{-i\, eB/(2\hbar c)\,(y_{1}x_{2}-y_{2}x_{1})}H_{B}\, e^{i\, eB/(2\hbar c)\,(y_{1}x_{2}-y_{2}x_{1})}.
\]

Shifting our focus to the covering space, suppose $\pmb{x}_{0},\pmb{x}\in\tilde{M}\cup\mathcal{A}\cup\mathcal{B}$
and $C\in\mathcal{A}\cup\mathcal{B}$ fulfill $\chi(\pmb{x}_{0},C)=\chi(\pmb{x},C)=1$,
i.e. $\pmb{x}_{0}$ and $\pmb{x}$ can be both connected with $C$
by a segment. In addition, suppose we are given $t_{0},t_{1}>0$.
In accordance with (\ref{eq:V_1sol}), (\ref{eq:V_U_aux}), we put
\begin{eqnarray*}
 &  & \hskip-2emV\!\left(\!\!\begin{array}{c}
\pmb{x},C,\pmb{x}_{0}\\
t_{1},t_{0}
\end{array}\!\!\right)\\
 &  & =\,\frac{\hbar}{i\mu}\left(\!\left(U-\frac{\omega_{c}\,(t_{1}+t_{0})}{2}+i\,(\Theta+\pi)\right)^{\!-1}-\left(U-\frac{\omega_{c}\,(t_{1}+t_{0})}{2}+i\,(\Theta-\pi)\right)^{\!-1}\right)
\end{eqnarray*}
where
\[
U=\ln\!\left(\frac{\dist(\pmb{x},C)\,\sinh(\omega_{c}t_{0}/2)}{\dist(\pmb{x}_{0},C)\,\sinh(\omega_{c}t_{1}/2)}\right)
\]
and $\Theta=\angle\,\pmb{x}_{0},C,\pmb{x}\in\mathbb{R}$ is the oriented
angle.

Denote by $\tilde{\sC}(\pmb{x},\pmb{x}_{0})$ the set of all piecewise
geodesic curves
\[
\gamma:\pmb{x}_{0}\to C_{1}\to\ldots\to C_{n}\to\pmb{x}
\]
with the inner vortices $C_{j}$, $1\leq j\leq n$, belonging to the
set of extreme points $\mathcal{A}\cup\mathcal{B}$. This means that
the equations $\chi(\pmb{x}_{0},C_{1})=\chi(C_{1},C_{2})=\ldots=\chi(C_{n},\pmb{x})=1$
hold. Let us denote by $|\gamma|=n\in\mathbb{N}$ the length of a
sequence $(C_{1},C_{2},\ldots,C_{n})$. To simplify the notation we
set everywhere where convenient $C_{0}=\pmb{x}_{0}$ and $C_{n+1}=\pmb{x}$.
Then formula (\ref{eq:ptilde_1sol_gen}) for the heat kernel $\tilde{p}_{t}(\pmb{x},\pmb{x}_{0})$
on the universal covering space $\tilde{M}$ generalizes to the model
with two AB fluxes as follows 
\begin{eqnarray}
\tilde{p}_{t}(\pmb{x},\pmb{x}_{0}) & = & \chi(\pmb{x},\pmb{x}_{0})\, p_{t}(\pmb{x},\pmb{x}_{0})\nonumber \\
 &  & +\,\sum_{\gamma\in\,\tilde{\sC}(x,x_{0})}\,\int_{0}^{\infty}\mathrm{d}t_{n}\int_{0}^{\infty}\mathrm{d}t_{n-1}\,\cdots\,\mathrm{\int_{0}^{\infty}d}t_{0}\,\delta(t_{n}+t_{n-1}+\ldots+t_{0}-t)\nonumber \\
\noalign{\smallskip} &  & \qquad\times\, p_{t_{n}}(\pmb{x},C_{n})\, p_{t_{n-1}}(C_{n},C_{n-1})\,\cdots\, p_{t_{0}}(C_{1},\pmb{x}_{0})\label{eq:ptilde_2sol}\\
\noalign{\smallskip} &  & \qquad\times\, V\!\left(\!\begin{array}{c}
\pmb{x},C_{n},C_{n-1}\\
t_{n},t_{n-1}
\end{array}\!\right)V\!\left(\!\begin{array}{c}
C_{n},C_{n-1},C_{n-2}\\
t_{n-1},t_{n-2}
\end{array}\!\right)\cdots\, V\!\left(\!\begin{array}{c}
C_{2},C_{1},\pmb{x}_{0}\\
t_{1},t_{0}
\end{array}\!\right)\!.\nonumber 
\end{eqnarray}
Here again, the symbol $p_{t}(\pmb{x},\pmb{y})$ is understood as
the lift of the heat kernel (\ref{eq:heat_ker_Cartes}) from the plane
to $\tilde{M}$ provided $\pmb{x}$ and $\pmb{y}$ can be connected
by a segment.

A detailed verification of (\ref{eq:ptilde_2sol}) in the particular
case when the background uniform magnetic field is absent can be found
in \cite{pk_ps_actapoly}. The general case comprising a uniform magnetic
field as well can be treated in the very same manner. We omit the
details here since this would be exceedingly lengthy.

\subsection{An application of the Schulman-Sunada formula}

One-dimensional unitary representations $\Lambda$ of $\Gamma$ are
determined by two numbers $\alpha$, $\beta$, $0\leq\alpha,\beta<1$,
such that 
\[
\Lambda(g_{a})=e^{2\pi i\alpha},\textrm{~}\Lambda(g_{b})=e^{2\pi i\beta}.
\]
Fix $\alpha,\beta\in(0,1)$ and, correspondingly, a representation
$\Lambda$. The values $\alpha$ and $\beta$ are proportional to
the strengths of the AB magnetic fluxes through the vortices $\pmb{a}$
and $\pmb{b}$, respectively (see (\ref{eq:PhiAB})). The heat kernel
$p_{t}^{\Lambda}(\pmb{x},\pmb{x}_{0})$ for a charged particle on
the plane pierced by these two AB fluxes and subjected to a uniform
magnetic field can again be derived with the aid the Schulman-Sunada
formula, now written in the form
\begin{equation}
p_{t}^{\Lambda}(\pmb{x},\pmb{x}_{0})=\sum_{g\in\Gamma}\Lambda(g^{-1})\,\tilde{p}_{t}(g\cdot\pmb{x},\pmb{x}_{0}).\label{eq:SuSch_2sol}
\end{equation}

Concerning the corresponding magnetic Schr\"odinger operator $H_{\Lambda}$,
it is convenient to pass to a unitarily equivalent formulation. Suppose
the vortices $\pmb{a}$, $\pmb{b}$ lie on the first axis and $\pmb{a}$
is located left to $\pmb{b}$. Let us cut the plane along two half-lines,
\[
L_{a}:=(-\infty,a_{1})\times\{0\}\textrm{~and~}L_{b}:=(b_{1},+\infty)\times\{0\}.
\]
Let $(r_{c},\theta_{c})$ be polar coordinates centered at a point
$\pmb{c}$, $\pmb{c}=\pmb{a},\pmb{b}$. Furthermore, the values $\theta_{a}=\pm\pi$
correspond to the two sides of the cut $L_{a}$, and similarly for
$\theta_{b}$ and $L_{b}$. The geometrical arrangement is depicted
on Figure~\ref{fig:geometry}.

The electromagnetic vector potential in the discussed model is the
sum of the vector potential $\pmb{A}_{B}$ given in (\ref{eq:avect_uni})
and the AB vector potential
\[
\pmb{A}_{AB}=-\frac{\hbar c}{e}\,(\alpha\nabla\theta_{a}+\beta\nabla\theta_{b}).
\]
The region $\mathcal{D}=\mathbb{R}^{2}\setminus(L_{a}\cup L_{b})$
is simply connected and therefore one can eliminate the vector potential
$\pmb{A}_{AB}$ by means of a gauge transform. This way one can pass
to a unitarily equivalent Hamiltonian $H_{\Lambda}'\simeq H_{\Lambda}$
acting as the differential operator $H_{B}$ in $L^{2}(\mathbb{R}^{2},\mbox{d}^{2}x)$,
see (\ref{eq:HB_Cartesian}), and whose domain is determined by the
boundary conditions along the cut, 
\begin{eqnarray}
 &  & \hspace{-3em}\psi(r_{a},\theta_{a}=\pi)=e^{2\pi\, i\,\alpha}\psi(r_{a},\theta_{a}=-\pi),\textrm{~}\partial_{\theta_{a}}\psi(r_{a},\theta_{a}=\pi)=e^{2\pi\, i\,\alpha}\,\partial_{\theta_{a}}\psi(r_{a},\theta_{a}=-\pi)\,,\nonumber \\
\noalign{\vskip-0.2em}\label{eq:bc}\\
\noalign{\vskip-0.6em} &  & \hspace{-3em}\psi(r_{b},\theta_{b}=\pi)=e^{2\pi\, i\,\beta}\psi(r_{b},\theta_{b}=-\pi),\textrm{~}\ \partial_{\theta_{b}}\psi(r_{b},\theta_{b}=\pi)=e^{2\pi\, i\,\beta}\,\partial_{\theta_{b}}\psi(r_{b},\theta_{b}=-\pi)\,.\nonumber 
\end{eqnarray}
In addition, one should impose the regular boundary condition at the
vortices, namely $\psi(\pmb{a})=\psi(\pmb{b})=0$.

One can embed $\mathcal{D}\subset\tilde{M}$ as a fundamental domain
of $\Gamma$. The heat kernel ${p'_{t}}^{\Lambda}(\pmb{x},\pmb{x}_{0})$
associated with ${H'}_{\Lambda}$ can be simply obtained as the restriction
to $\mathcal{D}$ of the heat kernel $p_{t}^{\Lambda}(\pmb{x},\pmb{x}_{0})$
associated with $H_{\Lambda}$ provided $p_{t}^{\Lambda}(\pmb{x},\pmb{x}_{0})$
is regarded as a $\Lambda$-equivariant function on $\tilde{M}\times\tilde{M}$.
In order to simplify the discussion of phases we shall assume that
either $\pmb{x}$, $\pmb{x}_{0}$ belong both to the upper half plane
or $\pmb{x}_{0}$ lies on the segment connecting $\pmb{a}$ and $\pmb{b}$.
In that case the segment $\overline{\pmb{x}_{0}\pmb{x}}$ intersects
none of the cuts $L_{a}$ and $L_{b}$.

Fix $t>0$. Let $\sC$ be the set of all finite alternating sequence
of points $\pmb{a}$ and $\pmb{b}$, i.e. $\overline{\gamma}\in\sC$
if $\overline{\gamma}=(\pmb{c}_{1},\ldots,\pmb{c}_{n})$ for some
$n\in\mathbb{N}$ and $\pmb{c}_{j}\in\{\pmb{a},\pmb{b}\}$, $\pmb{c}_{j}\neq\pmb{c}_{j+1}$.
Relate to $\overline{\gamma}$ a piecewise geodesic path in $M$,
namely
\begin{equation}
\pmb{x}_{0}\to\pmb{c}_{1}\to\ldots\to\pmb{c}_{n}\to\pmb{x}.\label{eq:path_M}
\end{equation}
Let us consider all possible piecewise geodesic paths in $\tilde{M}$,
\begin{equation}
\gamma:\pmb{x}_{0}\to C_{1}\to\ldots\to C_{n}\to g\cdot\pmb{x},\label{eq:path_Mtilde}
\end{equation}
with $C_{j}\in\mathcal{A}\cup\mathcal{B}$ and $g\in\Gamma$, covering
the path (\ref{eq:path_M}). Then $C_{j}\in\mathcal{A}$ if and only
if $\pmb{c}_{j}=\pmb{a}$, and $C_{j}\in\mathcal{B}$ if and only
if $\pmb{c}_{j}=\pmb{b}$. For $n\geq2$, denote the oriented angles
$\angle\,\pmb{x}_{0},\pmb{c}_{1},\pmb{c}_{2}=\theta_{0}$ and $\angle\,\pmb{c}_{n-1},\pmb{c}_{n},\pmb{x}=\theta$.
Then angles at the corresponding vortices in the path $\gamma$ take
the values $\angle\,\pmb{x}_{0},C_{1},C_{2}=\theta_{0}+2\pi k_{1}$,
$\angle\, C_{n-1},C_{n},g\cdot\pmb{x}=\theta+2\pi k_{n}$ and $\angle\, C_{j},C_{j+1},C_{j+2}=2\pi k_{j+1}$
for $1\leq j\leq n-2$ (if $n\geq3$), where $k_{1},\ldots,k_{n}$
are integers. For $n=1$, $\angle\,\pmb{x}_{0},C_{1},g\cdot\pmb{x}=\angle\,\pmb{x}_{0},c_{1},\pmb{x}+2\pi k_{1}$,
$k_{1}\in\mathbb{Z}$. Let us emphasize that any values $k_{1},\ldots,k_{n}\in\mathbb{Z}$
are possible, and these $n$-tuples of integers are in one-to-one
correspondence with the group elements $g$ occurring in (\ref{eq:path_Mtilde}).
The representation $\Lambda$ takes on $g$ the value 
\[
\Lambda(g)=\exp(2\pi i(k_{1}\sigma_{1}+\ldots+k_{n}\sigma_{n}))
\]
where $\sigma_{j}\in\{\alpha,\beta\}$ and $\sigma_{j}=\alpha$ if
$\pmb{x}_{j}=\pmb{a}$, and $\sigma_{j}=\beta$ if $\pmb{c}_{j}=\pmb{b}$.

Note that (\ref{eq:sum_aux1}) implies that
\begin{equation}
\sum_{k\in\mathbb{Z}}e^{-2\pi\mathrm{i}\alpha k}\left(\frac{1}{z+i\,(2k+1)\pi}-\frac{1}{z+i\,(2k-1)\pi}\right)=\frac{2}{i}\,\sin(\pi\alpha)\,\frac{e^{\alpha z}}{1+e^{z}}\label{eq:sum_k_V}
\end{equation}
holds for $0<\alpha<1$ and $z\in\mathbb{C}$. Observe also that the
expression
\[
p_{t_{n}}(\pmb{x},C_{n})\, p_{t_{n-1}}(C_{n},C_{n-1})\,\cdots\, p_{t_{0}}(C_{1},\pmb{x}_{0})
\]
in the integrand in (\ref{eq:ptilde_2sol}) in fact does not depend
on the integers $k_{1},\ldots,k_{n}$, see (\ref{eq:heat_extremA}).
Consequently, using (\ref{eq:sum_k_V}) one can carry out a partial
summation over $k_{1},\ldots,k_{n}$ in (\ref{eq:SuSch_2sol}), (\ref{eq:ptilde_2sol}).
This way the summation over $g\in\Gamma$ in (\ref{eq:SuSch_2sol})
reduces to a sum over $\overline{\gamma}\in\sC$.

Further we apply the following transformation of coordinates for a
given $t>0$ and $\overline{\gamma}\in\sC$. Denote again by $R$
the distance between $\pmb{a}$ and $\pmb{b}$. Put $r_{0}=|\pmb{x}_{0}-\pmb{c}_{1}|$,
$r_{j}=|\pmb{c}_{j}-\pmb{c}_{j+1}|=R$ for $0<j<n$, and $r_{n}=|\pmb{c}_{n}-\pmb{x}|$.
The transformation sends an $(n+1)$-tuple of positive numbers $(t_{0},t_{1},\ldots,t_{n})$
fulfilling
\[
t_{0}+t_{1}+\cdots+t_{n}=t
\]
to $u\in\mathbb{R}^{n}$,
\[
u_{j}=\ln\!\left(\frac{r_{j}\sinh(\omega_{c}t_{j-1}/2)}{r_{j-1}\sinh(\omega_{c}t_{j}/2)}\right)-\frac{\omega_{c}\,(t_{j-1}+t_{j})}{2},\ 1\leq j\leq n.
\]
Let us regard $t_{1},t_{2},\ldots,t_{n}>0$, obeying $0<t_{1}+t_{2}+\cdots+t_{n}<t$,
as independent variables and let $t_{0}=t-t_{1}-t_{2}-\cdots-t_{n}$.
The transformation is invertible, the inverse image of $u\in\mathbb{R}^{n}$
is
\[
t_{j}=\frac{1}{\omega_{c}}\ln\!\left(\frac{1+T_{j}(u)+e^{-\omega_{c}t}\left(T_{n}(u)-T_{j}(u)\right)}{1+T_{j-1}(u)+e^{-\omega_{c}t}\left(T_{n}(u)-T_{j-1}(u)\right)}\right),\text{ }\ j=1,2,\ldots,n.
\]
where we have put
\[
T_{j}(u)=\sum_{k=1}^{j}\frac{r_{k}}{r_{0}}\exp\!\left(-\sum_{\ell=1}^{k}u_{\ell}\right),\text{ }\ j=0,1,\ldots,n;
\]
particularly, $T_{0}(u)=0$.

By some routine manipulations one can verify that
\[
\mbox{d}u_{1}\mbox{d}u_{2}\ldots\mbox{d}u_{n}=\left(\frac{\omega_{c}}{2}\right)^{\! n}\frac{\sinh(\omega_{c}t/2)}{\sinh(\omega_{c}t_{0}/2)\sinh(\omega_{c}t_{1}/2)\cdots\sinh(\omega_{c}t_{n})}\,\mbox{d}t_{1}\mbox{d}t_{2}\ldots\mbox{d}t_{n}.
\]
Furthermore,
\[
\sum_{j=0}^{n}r_{j}^{\,2}\coth\!\left(\frac{\omega_{c}t_{j}}{2}\right)\!=\coth\!\left(\frac{\omega_{c}t}{2}\right)\sum_{j=0}^{n}r_{j}^{\,2}+\frac{2}{\sinh(\omega_{c}t/2)}\,\sum_{\substack{j,k=0\\
j<k
}
}^{n}r_{j}r_{k}\cosh\!\left(\frac{\omega_{c}t}{2}+\sum_{s=j+1}^{k}u_{s}\!\right)\!.
\]

Thus we arrive at the desired formula. Suppose $a_{1},b_{1}\in\mathbb{R\equiv\mathbb{R}}\times\{0\}\subset\mathbb{R}^{2}$,
$a_{1}<b_{1}$, $a_{2}=b_{2}=0$, $\alpha,\beta\in(0,1)$ and $\pmb{x}_{0},\pmb{x}\in\mathbb{R}_{+}^{2}$
(the upper half-plane), $t>0$. Put $R=b_{1}-a_{1}$. Then
\begin{equation}
{p'_{t}}^{\Lambda}(x,x_{0})={p'_{t}}^{\Lambda}(I)+{p'_{t}}^{\Lambda}(II)+{p'_{t}}^{\Lambda}(III)\label{eq:final_A}
\end{equation}
where
\begin{equation}
{p'_{t}}^{\Lambda}(I)=\frac{\mu\omega_{c}}{4\pi\hbar\sinh(\omega_{c}t/2)}\,\exp\!\left(-\frac{\mu\omega_{c}\,|\pmb{x}-\pmb{x}_{0}|^{2}}{4\hbar}\,\coth\!\left(\frac{\omega_{c}t}{2}\right)+i\,\frac{\mu\omega_{c}}{2\hbar}\,\pmb{x}\wedge\pmb{x}_{0}\right)\!,\label{eq:final_AI}
\end{equation}
\begin{eqnarray}
 &  & \hskip-5em{p'_{t}}^{\Lambda}(II)\,=\,-\,\frac{\mu\omega_{c}}{4\pi^{2}\hbar\sinh(\omega_{c}t/2)}\,\sum_{\pmb{c}\in\{\pmb{a},\pmb{b}\}}\sin(\pi\sigma)\nonumber \\
 &  & \qquad\qquad\times\,\exp\!\left(-\frac{\mu\omega_{c}\,(r^{2}+r_{0}^{\,2})}{4\hbar}\,\coth\!\left(\frac{\omega_{c}t}{2}\right)+i\,\frac{\mu\omega_{c}}{2\hbar}\,(\pmb{x}-\pmb{x}_{0})\wedge\pmb{c}\right)\label{eq:final_AII}\\
\noalign{\smallskip} &  & \qquad\qquad\times\,\int_{\mathbb{R}}\exp\!\left(-\frac{\mu\omega_{c}\, rr_{0}}{2\hbar\sinh(\omega_{c}t/2)}\,\cosh\!\left(\frac{\omega_{c}t}{2}+u\right)\!\right)\frac{e^{\sigma\,(u+i\theta_{1,0})}}{1+e^{u+i\theta_{1,0}}}\,\mbox{d}u,\nonumber 
\end{eqnarray}
\begin{eqnarray}
 &  & \hskip-2em{p'_{t}}^{\Lambda}(x,x_{0})\,=\,\sum_{\overline{\gamma}\in\sC,\, n\geq2}\frac{(-1)^{n}\mu\omega_{c}}{4\pi^{n+1}\hbar\sinh(\omega_{c}t/2)}\nonumber \\
 &  & \qquad\times\,\exp\!\left(-\frac{\mu\omega_{c}\,(r^{2}+(n-1)R^{2}+r_{0}^{\,2})}{4\hbar}\,\coth\!\left(\frac{\omega_{c}t}{2}\right)+i\,\frac{\mu\omega_{c}}{2\hbar}\,(\pmb{x}\wedge\pmb{c}_{n}+\pmb{c}_{1}\wedge\pmb{x}_{0})\right)\nonumber \\
\noalign{\smallskip} &  & \qquad\times\,\int_{\mathbb{R}^{n}}\exp\!\left(-\frac{\mu\omega_{c}}{2\hbar\sinh(\omega_{c}t/2)}\,\sum_{0\leq j<k\leq n}r_{k}r_{j}\cosh\!\left(\frac{\omega_{c}t}{2}+\sum_{s=j+1}^{k}u_{s}\right)\right)\label{eq:final_AIII}\\
\noalign{\smallskip} &  & \qquad\times\,\frac{\sin(\pi\sigma_{n})\, e^{\sigma_{n}\,(u_{n}+i\theta)}}{1+e^{u_{n}+i\theta}}\left(\prod_{j=2}^{n-1}\,\frac{\sin(\pi\sigma_{j})\, e^{\sigma_{j}u_{j}}}{1+e^{u_{j}}}\right)\frac{\sin(\pi\sigma_{1})\, e^{\sigma_{1}\,(u_{1}+i\theta_{0})}}{1+e^{u_{1}+i\theta_{0}}}\,\mbox{d}^{n}u.\nonumber 
\end{eqnarray}
Here we have put, in (\ref{eq:final_AII}), $r=|\pmb{x}-\pmb{c}|$,
$r_{0}=|\pmb{x}_{0}-\pmb{c}|$, $\theta_{1,0}=\angle\,\pmb{x}_{0},\pmb{c},\pmb{x}$
and $\sigma=\alpha$ (respectively, $\beta$) if $\pmb{c}=\pmb{a}$
(respectively, $\pmb{b}$), and in (\ref{eq:final_AIII}), $\overline{\gamma}=(\pmb{c}_{1},\pmb{c}_{2},\ldots,\pmb{c}_{n})$,
$r=r_{n}=|\pmb{x}-\pmb{c}_{n}|$, $r_{0}=|\pmb{x}_{0}-\pmb{c}_{1}|$,
$r_{j}=R$ for $1\leq j\leq n-1$, $\theta=\angle\,\pmb{c}_{n-1},\pmb{c}_{n},\pmb{x}$,
$\theta_{0}=\angle\,\pmb{x}_{0},\pmb{c}_{1},\pmb{c}_{2}$ and $\sigma_{j}=\alpha$
(respectively, $\beta$) if $\pmb{c}_{j}=\pmb{a}$ (respectively,
$\pmb{b}$).

\section{An approximate formula for a $2$-solenoid eigenfunction \label{sec:A_formula_eigenfunction}}

\subsection{An approximation of the heat kernel}

We are still using the notation introduced above, in particular, $(r_{a},\theta_{a})$
denote polar coordinates with respect to the center $\pmb{a}$, with
the equations $\theta_{a}=\pm\pi$ defining the two sides of the cut
$L_{a}$, and similarly for $(r_{b},\theta_{b})$ and $L_{b}$. Furthermore,
$R=\dist(\pmb{a},\pmb{b})$ and $\pmb{a},\pmb{b}\in\mathbb{R}\times\{0\}\subset\mathbb{R}^{2}$.
Assume that $\pmb{x}_{0}$ is an inner point of the segment $\overline{\pmb{a}\pmb{b}}$,
$\pmb{x}\in\mathbb{R}^{2}\setminus(L_{a}\cup L_{b})$. Note that,
with this setting, $r_{0a}+r_{0b}=R$, $\theta_{0a}=\theta_{0b}=0$
and $\pmb{x}_{0}\wedge\pmb{a}=\pmb{x}_{0}\wedge\pmb{b}=0$.

We start from formula (\ref{eq:final_A}) for the heat kernel. The
representation $\Lambda$ is determined by parameters $\alpha,\beta$,
and we assume that
\begin{equation}
0<\alpha<\beta<1.\label{eq:0_alpha_beta_1}
\end{equation}
For simplicity we omit the symbol $\Lambda$ in the notation and write
$H'$ instead of $H_{\Lambda}'$.

Suppose
\begin{equation}
D:=\frac{|e|BR^{2}}{\hbar c}=\frac{\mu\omega_{c}R^{2}}{\hbar}\gg1.\label{eq:D}
\end{equation}
In this asymptotic domain we adopt as an approximation of the heat
kernel a truncation of the series (\ref{eq:final_A}). In more detail,
we keep in (\ref{eq:final_AI}), (\ref{eq:final_AII}) and (\ref{eq:final_AIII})
only the terms up to the length $n=2$ of the sequence $\overline{\gamma}$.
Explicitly,

\begin{equation}
p'_{t}(x,x_{0})\simeq p'_{0}+p'_{a}+p'_{b}+p'_{ab}+p'_{ba},\label{eq:ptprime_approx}
\end{equation}
where
\[
p'_{0}=\frac{\mu\omega_{c}}{4\pi\hbar\sinh(\omega_{c}t/\hbar)}\,\exp\!\left(-\frac{\mu\omega_{c}\,|\pmb{x}-\pmb{x}_{0}|^{2}}{4\hbar}\coth\!\left(\frac{\omega_{c}t}{2}\right)+i\,\frac{\mu\omega_{c}}{2\hbar}\,\pmb{x}\wedge\pmb{x}_{0}\right)\!,
\]
\begin{eqnarray*}
p'_{a} & = & -\frac{\mu\omega_{c}\sin(\pi\alpha)}{4\pi^{2}\hbar\sinh(\omega_{c}t/2)}\,\exp\!\left(-\frac{\mu\omega_{c}}{4\hbar}\coth\!\left(\frac{\omega_{c}t}{2}\right)\,(r_{a}^{\,2}+r_{0a}^{\,2})+i\,\frac{\mu\omega_{c}}{2\hbar}\,\pmb{x}\wedge\pmb{a}\right)\\
 &  & \times\,\int_{-\infty}^{\infty}\exp\!\left(-\frac{\mu\omega_{c}\, r_{a}r_{0a}}{2\hbar\sinh(\omega_{c}t/2)}\cosh\!\left(\frac{\omega_{c}t}{2}+u\right)\right)\!\frac{e^{\alpha\,(u+i\theta_{a})}}{1+e^{u+i\theta_{a}}}\,\mbox{d}u,
\end{eqnarray*}
\begin{eqnarray*}
p'_{b} & = & -\frac{\mu\omega_{c}\sin(\pi\beta)}{4\pi^{2}\hbar\sinh(\omega_{c}t/2)}\,\exp\!\left(-\frac{\mu\omega_{c}}{4\hbar}\coth\!\left(\frac{\omega_{c}t}{2}\right)\,(r_{b}^{\,2}+r_{0b}^{\,2})+i\,\frac{\mu\omega_{c}}{2\hbar}\,\pmb{x}\wedge\pmb{b}\right)\\
 &  & \times\,\int_{-\infty}^{\infty}\exp\!\left(-\frac{\mu\omega_{c}\, r_{b}r_{0b}}{2\hbar\sinh(\omega_{c}t/2)}\cosh\!\left(\frac{\omega_{c}t}{2}+u\right)\right)\!\frac{e^{\beta\,(u+i\theta_{b})}}{1+e^{u+i\theta_{b}}}\,\mbox{d}u,
\end{eqnarray*}
\begin{eqnarray*}
 &  & p'_{ab}\,=\,\frac{\mu\omega_{c}\sin(\pi\alpha)\sin(\pi\beta)}{4\pi^{3}\sinh(\omega_{c}t/2)}\,\exp\!\left(-\frac{\mu\omega_{c}}{4\hbar}\coth\!\left(\frac{\omega_{c}t}{2}\right)\,(r_{b}^{\,2}+R^{2}+r_{0a}^{\,2})+i\,\frac{\mu\omega_{c}}{2\hbar}\,\pmb{x}\wedge\pmb{b}\right)\\
 &  & \quad\ \times\int_{-\infty}^{\infty}\int_{-\infty}^{\infty}\exp\!\Bigg(\!-\frac{\mu\omega_{c}}{2\hbar\sinh(\omega_{c}t/2)}\bigg(\! Rr_{0a}\cosh\!\left(\frac{\omega_{c}t}{2}+u_{1}\right)+Rr_{b}\cosh\!\left(\frac{\omega_{c}t}{2}+u_{2}\right)\\
 &  & \qquad\qquad\qquad\qquad\quad+\, r_{b}r_{0a}\cosh\!\left(\frac{\omega_{c}t}{2}+u_{1}+u_{2}\right)\!\bigg)\!\Bigg)\frac{e^{\alpha u_{1}+\beta\,(u_{2}+i\theta_{b})}}{(1+e^{u_{1}})(1+e^{u_{2}+i\theta_{b}})}\,\mbox{d}u_{1}\mbox{d}u_{2},
\end{eqnarray*}
\begin{eqnarray*}
 &  & p'_{ab}\,=\,\frac{\mu\omega_{c}\sin(\pi\alpha)\sin(\pi\beta)}{4\pi^{3}\hbar\sinh(\omega_{c}t/2)}\,\exp\!\left(-\frac{\mu\omega_{c}}{4\hbar}\coth\!\left(\frac{\omega_{c}t}{2}\right)\,(r_{a}^{\,2}+R^{2}+r_{0b}^{\,2})+i\,\frac{\mu\omega_{c}}{2\hbar}\,\pmb{x}\wedge\pmb{a}\right)\\
 &  & \quad\times\int_{-\infty}^{\infty}\int_{-\infty}^{\infty}\exp\!\Bigg(\!-\frac{\mu\omega_{c}}{2\sinh(\omega_{c}t/2)}\bigg(Rr_{0b}\cosh\!\left(\frac{\omega_{c}t}{2}+u_{1}\right)+Rr_{a}\cosh\!\left(\frac{\omega_{c}t}{2}+u_{2}\right)\\
 &  & \qquad\qquad\qquad\qquad\quad+\, r_{a}r_{0b}\cosh\!\left(\frac{\omega_{c}t}{2}+u_{1}+u_{2}\right)\!\bigg)\!\Bigg)\frac{e^{\beta u_{1}+\alpha\,(u_{2}+i\theta_{a})}}{(1+e^{u_{1}})(1+e^{u_{2}+i\theta_{a}})}\,\mbox{d}u_{1}\mbox{d}u_{2}.
\end{eqnarray*}

\subsection{Extracting an eigenfunction from the heat kernel}

We have $p'_{t}(\pmb{x},\pmb{x}_{0})=\langle\pmb{x}|\exp(-H't/\hbar)|\pmb{x}_{0}\rangle$
and
\[
e^{-H't/\hbar}=\sum_{\lambda\in\spec(H')}e^{-\lambda t/\hbar}P_{\lambda}
\]
where $P_{\lambda}$ is the eigenprojection corresponding to an eigenvalue
$\lambda$. This means that looking at the asymptotic expansion of
the heat kernel for large time one can, in principle, extract from
it all information about the spectral properties of the Hamiltonian.
Assuming (\ref{eq:0_alpha_beta_1}) and (\ref{eq:D}) our goal is
to derive an approximate formula for the lowest eigenvalue above the
first Landau level. This is to say that we are interested in the unique
simple eigenvalue located near
\begin{equation}
E_{1}:=\left(\alpha+\frac{1}{2}\right)\!\hbar\omega_{c}\label{eq:E1}
\end{equation}
 as well as in a corresponding normalized eigenfunction which we denote
$E_{2}$ and $\psi_{2}$, respectively.

Thus we are lead to a discussion of the asymptotic behavior of the
approximation (\ref{eq:ptprime_approx}) as $t\to+\infty$. It is
immediately seen that
\begin{equation}
p'_{0}=\frac{\mu\omega_{c}}{4\pi\hbar}\,\exp\!\left(-\frac{\mu\omega_{c}}{4\hbar}\,|\pmb{x}-\pmb{x}_{0}|^{2}+i\,\frac{\mu\omega_{c}}{2\hbar}\,\pmb{x}\wedge\pmb{x}_{0}\right)\! e^{-\omega_{c}t/2}+O(e^{-3\omega_{c}t/2}).\label{eq:pasympt_0}
\end{equation}
To treat $p'_{a}$ we apply (\ref{eq:asympt_single}) where we set
$X=\mu\omega_{c}\, r_{a}r_{0a}/(2\hbar)$. Hence
\begin{eqnarray}
p'_{a} & = & -\frac{\mu\omega_{c}\sin(\pi\alpha)}{2\pi^{2}\hbar}\,\exp\!\left(-\frac{\mu\omega_{c}}{4\hbar}\,(r_{a}^{\,2}+r_{0a}^{\,2})+i\,\frac{\mu\omega_{c}}{2\hbar}\,\pmb{x}\wedge\pmb{a}\right)\nonumber \\
\noalign{\smallskip} &  & \qquad\ \ \times\,\int_{-\infty}^{\infty}\exp\!\left(-\frac{\mu\omega_{c}\, r_{a}r_{0a}}{2\hbar}\, e^{u}\right)\!\frac{e^{\alpha\,(u+i\theta_{a})}}{1+e^{u+i\theta_{a}}}\,\mbox{d}u\,\times e^{-\omega_{c}t/2}\label{eq:pa_asympt}\\
\noalign{\smallskip} &  & +\,\frac{\mu\omega_{c}}{2\pi\Gamma(1+\alpha)\hbar}\,\exp\!\left(-\frac{\mu\omega_{c}}{4\hbar}\,(r_{a}^{\,2}+r_{0a}^{\,2})+i\,\frac{\mu\omega_{c}}{2\hbar}\,\pmb{x}\wedge\pmb{a}\right)\!\left(\frac{\mu\omega_{c}\, r_{a}r_{0a}}{2\hbar}\right)^{\!\alpha}e^{i\alpha\theta_{a}}\nonumber \\
\noalign{\smallskip} &  & \qquad\ \ \times\, e^{-(\alpha+1/2)\omega_{c}t/2}\,+\, O(e^{-3\omega_{c}t/2}).\nonumber 
\end{eqnarray}
An analogous formula for $p'_{b}$ reads
\begin{eqnarray}
p'_{b} & = & -\frac{\mu\omega_{c}\sin(\pi\beta)}{2\pi^{2}\hbar}\,\exp\!\left(-\frac{\mu\omega_{c}}{4\hbar}\,(r_{b}^{\,2}+r_{0b}^{\,2})+i\,\frac{\mu\omega_{c}}{2\hbar}\,\pmb{x}\wedge\pmb{b}\right)\label{eq:pb_asympt}\\
\noalign{\smallskip} &  & \ \times\int_{-\infty}^{\infty}\exp\!\left(-\frac{\mu\omega_{c}\, r_{b}r_{0b}}{2\hbar}\, e^{u}\right)\!\frac{e^{\beta\,(u+i\theta_{b})}}{1+e^{u+i\theta_{b}}}\,\mbox{d}u\,\times e^{-\omega_{c}t/2}+\, O(e^{-(\beta+1/2)\omega_{c}t}).\nonumber 
\end{eqnarray}

To treat $p'_{ab}$ we apply the asymptotic formula (\ref{eq:intV_asympt_tot})
in which we set $X=\mu\omega_{c}\, Rr_{0a}/(2\hbar)$, $Y=\mu\omega_{c}\, Rr_{b}/(2\hbar)$,
$Z=\mu\omega_{c}\, r_{0a}r_{b}/(2\hbar)$, thus obtaining
\begin{eqnarray}
p'_{ab} & = & \frac{\mu\omega_{c}\sin(\pi\alpha)\sin(\pi\beta)}{2\pi^{3}\hbar}\,\exp\!\left(-\frac{\mu\omega_{c}}{4\hbar}\,(r_{b}^{\,2}+R^{2}+r_{0a}^{\,2})+i\,\frac{\mu\omega_{c}}{2\hbar}\,\pmb{x}\wedge\pmb{b}\right)\nonumber \\
\noalign{\smallskip} &  & \quad\times\int_{-\infty}^{\infty}\int_{-\infty}^{\infty}\exp\!\bigg(\!-\frac{\mu\omega_{c}}{2\hbar}\big(Rr_{0a}e^{u_{1}}+Rr_{b}e^{u_{2}}+r_{0a}r_{b}e^{u_{1}+u_{2}}\big)\!\bigg)\label{eq:pab_asympt}\\
 &  & \qquad\qquad\qquad\ \times\,\frac{e^{\alpha u_{1}+\beta\,(u_{2}+i\theta_{b})}}{(1+e^{u_{1}})(1+e^{u_{2}+i\theta_{b}})}\,\mbox{d}u_{1}\mbox{d}u_{2}\,\times e^{-\omega_{c}t/2}\nonumber \\
 &  & -\frac{\mu\omega_{c}\sin(\pi\beta)}{2\pi^{2}\Gamma(1+\alpha)\hbar}\exp\!\left(-\frac{\mu\omega_{c}}{4\hbar}\,(r_{b}^{\,2}+R^{2}+r_{0a}^{\,2})+i\,\frac{\mu\omega_{c}}{2\hbar}\,\pmb{x}\wedge\pmb{b}\right)\!\!\left(\frac{\mu\omega_{c}Rr_{0a}}{2\hbar}\right)^{\!\alpha}\nonumber \\
\noalign{\smallskip} &  & \quad\times\!\int_{-\infty}^{\infty}\left(1+\frac{r_{b}}{R}e^{-u}\right)^{\!\alpha}\exp\!\left(-\frac{\mu\omega_{c}Rr_{b}}{2\hbar}\, e^{u}\right)\frac{e^{\beta(u+i\theta_{b})}}{1+e^{u+i\theta_{b}}}\,\mbox{d}u\,\times\, e^{-(\alpha+1/2)\omega_{c}t}\nonumber \\
\noalign{\smallskip} &  & +\, O(e^{-(\beta+1/2)\omega_{c}t}).\nonumber 
\end{eqnarray}
An analogous formula for $p'_{ba}$ reads
\begin{eqnarray}
p'_{ba} & = & \frac{\mu\omega_{c}\sin(\pi\alpha)\sin(\pi\beta)}{2\pi^{3}\hbar}\,\exp\!\left(-\frac{\mu\omega_{c}}{4\hbar}\,(r_{a}^{\,2}+R^{2}+r_{0b}^{\,2})+i\,\frac{\mu\omega_{c}}{2\hbar}\,\pmb{x}\wedge\pmb{a}\right)\nonumber \\
\noalign{\smallskip} &  & \quad\times\int_{-\infty}^{\infty}\int_{-\infty}^{\infty}\exp\!\bigg(\!-\frac{\mu\omega_{c}}{2\hbar}\big(Rr_{0b}e^{u_{1}}+Rr_{a}e^{u_{2}}+r_{0b}r_{a}e^{u_{1}+u_{2}}\big)\!\bigg)\label{eq:pasympt_2ba}\\
 &  & \qquad\qquad\quad\ \times\,\frac{e^{\beta u_{1}+\alpha\,(u_{2}+i\theta_{b})}}{(1+e^{u_{1}})(1+e^{u_{2}+i\theta_{a}})}\,\mbox{d}u_{1}\mbox{d}u_{2}\,\times e^{-\omega_{c}t/2}\nonumber \\
 &  & -\frac{\mu\omega_{c}\sin(\pi\beta)}{2\pi^{2}\Gamma(1+\alpha)\hbar}\exp\!\left(-\frac{\mu\omega_{c}}{4\hbar}\,(r_{a}^{\,2}+R^{2}+r_{0b}^{\,2})+i\,\frac{\mu\omega_{c}}{2\hbar}\,\pmb{x}\wedge\pmb{a}\right)\!\!\left(\frac{BRr_{a}}{2}\right)^{\!\alpha}e^{i\alpha\theta_{a}}\nonumber \\
\noalign{\smallskip} &  & \quad\times\int_{-\infty}^{\infty}\left(1+\frac{r_{0b}}{R}e^{-u}\right)^{\!\alpha}\exp\!\left(-\frac{\mu\omega_{c}Rr_{0b}}{2\hbar}\, e^{u}\right)\frac{e^{\beta u}}{1+e^{u}}\,\mbox{d}u\,\times\, e^{-(\alpha+1/2)\omega_{c}t}\nonumber \\
\noalign{\smallskip} &  & +\, O(e^{-(\beta+1/2)\omega_{c}t}).\nonumber 
\end{eqnarray}

Now we are in the position to derive an approximation of the eigenfunction
$\psi_{2}$. As far as the eigenvalue $E_{2}$ is concerned, it turns
out, unfortunately, that the approximation (\ref{eq:ptprime_approx})
is incapable to directly distinguish between $E_{2}$ and $E_{1}$.
As pointed out in the introduction, these eigenvalues approach each
other extremely rapidly as $D$ becomes large. However, having at
our disposal an approximate formula for the eigenfunction we shall
be able to derive, a posteriori in Section~\ref{sec:DeltaE}, a perturbative
formula for the difference between $E_{2}$ and $E_{1}$.

Since $E_{2}$ is simple we have
\[
\langle\pmb{x}|P_{E_{2}}|\pmb{x}_{0}\rangle=\psi_{2}(\pmb{x})\,\overline{\psi_{2}(\pmb{x}_{0})}\,.
\]
From the asymptotic expansion (\ref{eq:palpha_1sol_asympt}) in Subsection~\ref{subsec:one_ABsolen},
if combined with (\ref{eq:pt_transform}), it can be seen that the
normalized one-solenoid eigenfunction $\psi_{1}$ corresponding to
$E_{1}$ fulfills
\begin{eqnarray*}
 &  & \psi_{1}(\pmb{x})\,\overline{\psi_{1}(\pmb{x}_{0})}\\
 &  & =\,\frac{\mu\omega_{c}}{2\pi\Gamma(1+\alpha)\hbar}\,\exp\!\left(-\frac{\mu\omega_{c}}{4\hbar}\,(r_{a}^{\,2}+r_{0a}^{\,2})+i\,\frac{\mu\omega_{c}}{2\hbar}\,(\pmb{x}-\pmb{x}_{0})\wedge\pmb{a}\right)\!\left(\frac{\mu\omega_{c}\, r_{a}r_{0a}}{2\hbar}\right)^{\!\alpha}e^{i\alpha(\theta_{a}-\theta_{0a})}.
\end{eqnarray*}
In our arrangement, $\theta_{0a}=0$ and $\pmb{x}_{0}\wedge\pmb{a}=0$,
and therefore
\begin{equation}
\psi_{1}(\pmb{x})=\left(\frac{\mu\omega_{c}}{2\pi\Gamma(1+\alpha)\hbar}\right)^{\!1/2}\exp\!\left(-\frac{\mu\omega_{c}}{4\hbar}\, r_{a}^{\,2}+i\,\frac{\mu\omega_{c}}{2\hbar}\,\pmb{x}\wedge\pmb{a}\right)\!\left(\sqrt{\frac{\mu\omega_{c}}{2\hbar}}\, r_{a}\right)^{\!\alpha}e^{i\alpha\theta_{a}}.\label{eq:psi1sol}
\end{equation}
Inspecting relations (\ref{eq:pasympt_0}) through (\ref{eq:pasympt_2ba})
and collecting the coefficients standing at $e^{-(\alpha+1/2)\omega_{c}t}$
we arrive at the expression
\[
\psi_{2}(\pmb{x})\,\overline{\psi_{2}(\pmb{x}_{0})}\simeq\psi_{1}(\pmb{x})\,\overline{\psi_{1}(\pmb{x}_{0})}+\psi_{1}(\pmb{x})\,\overline{\varphi(\pmb{x}_{0})}+\varphi(\pmb{x})\,\overline{\psi_{1}(\pmb{x}_{0})}
\]
where
\begin{eqnarray}
\varphi(\pmb{x}) & = & -\sqrt{\frac{\mu\omega_{c}}{2\pi\Gamma(1+\alpha)\hbar}}\,\frac{\sin(\pi\beta)}{\pi}\,\exp\!\left(-\frac{\mu\omega_{c}}{4\hbar}\,(R^{2}+r_{b}^{\,2})+i\,\frac{\mu\omega_{c}}{2\hbar}\,\pmb{x}\wedge\pmb{b}\right)\nonumber \\
 &  & \times\int_{-\infty}^{\infty}\left(\sqrt{\frac{\mu\omega_{c}}{2\hbar}}\,(R+r_{b}e^{-u})\right)^{\!\alpha}\exp\!\left(-\frac{\mu\omega_{c}Rr_{b}}{2\hbar}\, e^{u}\right)\frac{e^{\beta(u+i\theta_{b})}}{1+e^{u+i\theta_{b}}}\,\mbox{d}u.\label{eq:W}
\end{eqnarray}
Recalling (\ref{eq:D}) and neglecting the term $\varphi(\pmb{x})\,\overline{\varphi(\pmb{x}_{0})}$
we adopt as an asymptotic approximation of the normalized eigenfunction
\begin{equation}
\psi_{2}(\pmb{x})\simeq\tilde{\psi}_{2}(\pmb{x}):=\psi_{1}(\pmb{x})+\varphi(\pmb{x}).\label{eq:psi2_W}
\end{equation}
Let us note that, if working with polar coordinates, we have $\pmb{x}\wedge\pmb{a}=-a_{1}r_{a}\sin(\theta_{a})$
and $\pmb{x}\wedge\pmb{b}=b_{1}r_{b}\sin(\theta_{b})$ in (\ref{eq:psi1sol})
and (\ref{eq:W}), respectively ($\pmb{a}=(a_{1},0)$, $\pmb{b}=(b_{1},0)$).

\begin{remark*} The wave function $\tilde{\psi}_{2}(\pmb{x})$ vanishes
at $\pmb{x}=\pmb{b}$. In fact, if $\pmb{x}=\pmb{b}$ then $r_{a}=R$,
$r_{b}=0$, $\theta_{a}=0$, $\theta_{b}$ is not defined and $\pmb{x}\wedge\pmb{a}=\pmb{x}\wedge\pmb{b}=0$.
A simple computation yields
\[
\tilde{\psi}_{2}(\pmb{b})\!\left(\frac{\mu\omega_{c}}{2\pi\Gamma(1+\alpha)\hbar}\right)^{\!-1/2}\exp\!\left(\frac{\mu\omega_{c}}{4\hbar}\, R^{2}\right)\!\!\left(\!\sqrt{\frac{\mu\omega_{c}}{2\hbar}}\, R\right)^{\!-\alpha}=1-\frac{\sin(\pi\beta)}{\pi}\int_{-\infty}^{\infty}\frac{e^{\beta(u+i\theta_{b})}}{1+e^{u+i\theta_{b}}}\,\mbox{d}u,
\]
and
\[
\int_{-\infty}^{\infty}\frac{e^{\beta(u+i\theta_{b})}}{1+e^{u+i\theta_{b}}}\,\mbox{d}u=\int_{-\infty}^{\infty}\frac{e^{\beta u}}{1+e^{u}}\,\mbox{d}u=\frac{\pi}{\sin(\pi\beta)}\,.
\]
\end{remark*}

\subsection{Another form of the approximate eigenfunction}

We are going to show that formula (\ref{eq:W}) can be simplified
provided (\ref{eq:0_alpha_beta_1}) is assumed. Here and in what follows,
\[
a_{1}=0,\ b_{1}=R.
\]
In the region determined by $r_{b}<R$ and $|\theta_{b}|<\pi/2$,
\begin{eqnarray*}
 &  & \int_{-\infty}^{\infty}\left(1+\frac{r_{b}}{R}\, e^{-u}\right)^{\!\alpha}\exp\!\left(-\frac{\mu\omega_{c}Rr_{b}}{2\hbar}\, e^{u}\right)\!\frac{e^{\beta(u+i\theta_{b})}}{1+e^{u+i\theta_{b}}}\,\mbox{d}u\\
 &  & =\int_{-\infty}^{\infty}\left(e^{u}+\frac{r_{b}}{R}\, e^{i\theta_{b}}\right)^{\!\alpha}\exp\!\left(-\frac{\mu\omega_{c}Rr_{b}}{2\hbar}\, e^{u-i\theta_{b}}\right)\!\frac{e^{(\beta-\alpha)u}}{1+e^{u}}\,\mbox{d}u,
\end{eqnarray*}
and, in view of (\ref{eq:int_U}), we get
\begin{eqnarray}
\varphi(\pmb{x}) & = & -\sqrt{\frac{\mu\omega_{c}}{2\pi\Gamma(\alpha+1)\hbar}}\left(\sqrt{\frac{\mu\omega_{c}}{2\hbar}}R\right)^{\alpha}\exp\!\left(-\frac{\mu\omega_{c}}{4\hbar}\left(R^{2}+r_{b}^{\,2}\right)+\frac{\mu\omega_{c}Rr_{b}}{2\hbar}\,\cos(\theta_{b})\right)\nonumber \\
 &  & \times\Bigg((1-z)^{\alpha}-\frac{\Gamma(\beta-\alpha)\,_{2}F_{1}(1,\beta-\alpha;\beta+1;z)\, z^{\beta}}{\Gamma(-\alpha)\Gamma(\beta+1)}\label{eq:phi_U2}\\
 &  & \qquad-\frac{\sin(\pi\beta)\Gamma(\beta-\alpha)}{\pi}\left(\frac{D\bar{z}}{2}\right)^{\!1-\beta}\int_{0}^{1}U\!\left(-\alpha,1-\beta,\frac{Dz\bar{z}t}{2}\right)e^{-D\bar{z}t/2}t^{-\beta}\,\mbox{d}t\Bigg)\nonumber 
\end{eqnarray}
where
\begin{equation}
z=\frac{r_{b}}{R}\, e^{i\theta_{b}},\ \bar{z}=\frac{r_{b}}{R}\, e^{-i\theta_{b}}.\label{eq:phi_c_z}
\end{equation}

One observes that $\varphi(\pmb{x})$, as given in (\ref{eq:W}),
is real analytic on $\mathbb{R}^{2}\setminus L_{b}$. This is in agreement
with the theorem about analytic elliptic regularity ensuring that
solutions of elliptic partial differential equations with real analytic
coefficients on an open set in $\mathbb{R}^{n}$ are themselves real
analytic on this set; see, for instance, \cite[Chp. VII]{John} or
\cite[\S 6C]{Folland}. Formula (\ref{eq:phi_U2}) has been derived
for a smaller region but referring to the uniqueness of the analytic
continuation we conclude that (\ref{eq:phi_U2}) is even valid as
long as $r_{b}<R$ and $|\theta_{b}|<\pi$.

Recalling (\ref{eq:psi1sol}) (with $\pmb{a}=\pmb{0}$) and observing
that
\begin{equation}
r_{a}e^{i\theta_{a}}=R-r_{b}e^{i\theta_{b}},\label{eq:polar_ab}
\end{equation}
i.e. $1-z=(r_{a}/R)\, e^{i\theta_{a}}$ and $R^{2}+r_{b}^{\,2}-2Rr_{b}\cos(\theta_{b})=r_{a}^{\,2}$,
we find that the approximate 2-solenoid eigenfunction equals
\begin{eqnarray}
\tilde{\psi}_{2}(\pmb{x}) & = & \psi_{1}(\pmb{x})+\varphi(\pmb{x})\nonumber \\
\noalign{\smallskip} & = & \sqrt{\frac{B}{2\pi\Gamma(\alpha+1)}}\left(\sqrt{\frac{B}{2}}R\right)^{\alpha}\exp\!\left(-\frac{B}{4}\left(R^{2}+r_{b}^{\,2}\right)+\frac{BR}{2}\, r_{b}\cos(\theta_{b})\right)\nonumber \\
 &  & \times\Bigg(\frac{\Gamma(\beta-\alpha)}{\Gamma(-\alpha)\Gamma(\beta+1)}\,\,_{2}F_{1}(1,\beta-\alpha;\beta+1;z)\, z^{\beta}\label{eq:psi2t_Lb}\\
 &  & \qquad+\,\frac{\sin(\pi\beta)\Gamma(\beta-\alpha)}{\pi}\,\left(\frac{D\bar{z}}{2}\right)^{\!1-\beta}\int_{0}^{1}U\!\left(-\alpha,1-\beta,\frac{Dz\bar{z}t}{2}\right)\, e^{-D\bar{z}t/2}t^{-\beta}\,\mbox{d}t\Bigg).\nonumber 
\end{eqnarray}

Note that from (\ref{eq:psi2t_Lb}) it is apparent that $\tilde{\psi}_{2}(\pmb{x})$
obeys the boundary conditions on the cut $L_{b}$, see (\ref{eq:bc}).

\subsection{The approximate eigenfunction is an exact solution of an eigenvalue
equation \label{subsec:approx_exact_ef}}

Still assuming (\ref{eq:0_alpha_beta_1}). Recall (\ref{eq:phi_U2})
and (\ref{eq:psi2t_Lb}). In this subsection we simplify the notation
by omitting the subscript $b$, and write $r=r_{b}$, $\theta=\theta_{b}$.
Moreover, it is convenient to use complex coordinates (\ref{eq:phi_c_z}).

Let us introduce a differential operator $H_{B}^{\text{max}}$ on
the open set $\mathbb{R}^{2}\setminus L_{a}\cup L_{b}$, formally
given by the same differential expression as $H_{B}$ in (\ref{eq:HB_Cartesian})
but with the maximal admissible domain as a differential operator
on the specified region. In more detail, $\psi\in L^{2}(\mathbb{R}^{2})$
belongs to $\Dom H_{B}^{\text{max}}$ if and only if $H_{B}^{\text{max}}\psi$
computed in the distributional sense on $\mathbb{R}^{2}\setminus L_{a}\cup L_{b}$
belongs as well to $L^{2}(\mathbb{R}^{2})$. In particular this means
that no boundary conditions are imposed on the cut $L_{a}\cup L_{b}$.
Another and mathematically rigorous formulation says that $H_{B}^{\text{max}}$
is the adjoint operator to the symmetric operator $\dot{H}_{B}$ which
is also formally given by the same differential expression (\ref{eq:HB_Cartesian})
but whose domain equals $C_{0}^{\infty}(\mathbb{R}^{2}\setminus L_{a}\cup L_{b})$
($C_{0}^{\infty}$ standing for compactly supported smooth functions
on the indicated region).

We claim that $\varphi(\pmb{x})$ is an exact eigenfunction of $H_{B}^{\text{max}}$
corresponding to the eigenvalue $E_{1}$, see (\ref{eq:E1}). Let
us note that, if expressed in polar coordinates $(r,\theta)$ (which
are centered at $\pmb{b}=(R,0)$), $H_{B}^{\text{max}}$ acts as the
differential operator
\begin{eqnarray*}
 &  & -\frac{\hbar^{2}}{2\mu}\Bigg(\left(-\cos(\theta)\frac{\partial}{\partial r}+\frac{\sin(\theta)}{r}\frac{\partial}{\partial\theta}-\frac{ieB}{2\hbar c}\, r\sin(\theta)\right)^{\!2}\\
 &  & \qquad\quad+\left(\sin(\theta)\frac{\partial}{\partial r}+\frac{\cos(\theta)}{r}\frac{\partial}{\partial\theta}+\frac{ieBR}{2\hbar c}-\frac{ieB}{2\hbar c}\, r\cos(\theta)\right)^{\!2}\Bigg).
\end{eqnarray*}

For convenience, let us move the origin of coordinates from $\pmb{a}=(0,0)$
to $\pmb{b}=(R,0)$. This means a change of the vector potential whose
original form was (\ref{eq:avect_uni}). This is done by means of
a gauge transformation, and the operator $H_{B}^{\,\text{max}}$ is
transformed correspondingly,
\[
H_{B}^{\text{max}}\to\exp\!\left(\frac{ieBR}{2\hbar c}\, r\sin(\theta)\right)H_{B}^{\text{max}}\exp\!\left(-\frac{ieBR}{2\hbar c}\, r\sin(\theta)\right).
\]
The resulting differential operator reads
\[
-\frac{\hbar^{2}}{2\mu}\Bigg(\frac{\partial^{2}}{\partial r^{2}}+\frac{1}{r}\frac{\partial}{\partial r}+\frac{1}{r^{2}}\!\left(\frac{\partial}{\partial\theta}-\frac{ieBr^{2}}{2\hbar c}\right)^{\!2}\Bigg).
\]
Recalling (\ref{eq:D}) and using complex coordinates, the differential
operator takes the form $\hbar^{2}/(2\mu R^{2})H_{I}$ where
\[
H_{I}=-4\,\frac{\partial^{2}}{\partial z\partial\bar{z}}+\frac{D^{2}}{4}\, z\bar{z}+D\left(z\,\frac{\partial}{\partial z}-\bar{z}\,\frac{\partial}{\partial\bar{z}}\right)\!,
\]
and the eigenfunction reads, up to a constant multiplier,
\begin{eqnarray*}
f_{I}\left(z,\bar{z}\right) & = & \exp\!\left(-\frac{Dz\bar{z}}{4}+\frac{D\bar{z}}{2}\right)\\
 &  & \times\Bigg(\frac{\Gamma(\beta-\alpha)\,_{2}F_{1}(1,\beta-\alpha;\beta+1;z)\, z^{\beta}}{\Gamma(-\alpha)\Gamma(1+\beta)}\\
 &  & \quad\ +\,\frac{\sin(\pi\beta)\Gamma(\beta-\alpha)}{\pi}\!\left(\frac{D\bar{z}}{2}\right)^{\!1-\beta}\int_{0}^{1}U\!\left(-\alpha,1-\beta,\frac{Dz\bar{z}t}{2}\right)\! e^{-D\bar{z}t/2}\, t^{-\beta}\,\mbox{d}t\Bigg)\!.
\end{eqnarray*}

We have to verify that $H_{I}f_{I}=(1+2\alpha)Df_{I}$. To this end,
let us apply yet another gauge transformation

\[
\exp\!\left(\frac{Dz\bar{z}}{4}-\frac{D\bar{z}}{2}\right)H_{I}\exp\!\left(-\frac{Dz\bar{z}}{4}+\frac{D\bar{z}}{2}\right)=2H_{II}+D
\]
where
\[
H_{II}=-2\,\frac{\partial^{2}}{\partial z\partial\bar{z}}+D(z-1)\,\frac{\partial}{\partial z}.
\]
The transformed eigenfunction is simply
\begin{eqnarray*}
f_{II}\left(z,\bar{z}\right) & = & \frac{\Gamma(\beta-\alpha)\,_{2}F_{1}(1,\beta-\alpha;\beta+1;z)\, z^{\beta}}{\Gamma(-\alpha)\Gamma(1+\beta)}\\
 &  & +\,\frac{\sin(\pi\beta)\Gamma(\beta-\alpha)}{\pi}\left(\frac{D\bar{z}}{2}\right)^{1-\beta}\int_{0}^{1}U\!\left(-\alpha,1-\beta,\frac{D}{2}z\bar{z}t\right)e^{-D\bar{z}t/2}\, t^{-\beta}\,\mbox{d}t,
\end{eqnarray*}
and it should hold $H_{II}f_{II}=\alpha Df_{II}$.

Finally, after rescaling $f_{I}(z,\bar{z})=g\left(Dz,D\bar{z}\right)$
we obtain the equation
\[
Hg=\frac{\alpha}{D}\, g,\ \text{with\ }\ H=-2\,\frac{\partial^{2}}{\partial z\partial\bar{z}}-\left(1-\frac{z}{D}\right)\frac{\partial}{\partial z}
\]
and
\begin{eqnarray*}
g(z,\bar{z}) & = & \frac{\Gamma(\beta-\alpha)}{\Gamma(-\alpha)\Gamma(1+\beta)}\,\,_{2}F_{1}\!\left(1,\beta-\alpha;\beta+1;\frac{z}{D}\right)z^{\beta}D^{-\beta}\\
 &  & +\,\frac{\sin(\pi\beta)\Gamma(\beta-\alpha)}{\pi}\int_{0}^{\bar{z}/2}U\!\left(-\alpha,1-\beta,\frac{zt}{D}\right)e^{-t}t^{-\beta}\,\mbox{d}t.
\end{eqnarray*}
Recall (\ref{eq:U}). We can write
\[
g\left(z,\bar{z}\right)=\frac{1}{\Gamma(1-\beta)}\, g_{1}\left(z,\bar{z}\right)+\frac{\Gamma(\beta-\alpha)D^{-\beta}}{\Gamma(-\alpha)\Gamma(1+\beta)}\, g_{2}\left(z,\bar{z}\right)
\]
where
\[
g_{1}\left(z,\bar{z}\right)=\int_{0}^{\bar{z}/2}\,_{1}F_{1}\!\left(-\alpha;1-\beta;\frac{zt}{D}\right)e^{-t}t^{-\beta}\,\mbox{d}t
\]
and
\[
g_{2}\left(z,\bar{z}\right)=\,_{2}F_{1}\!\left(1,\beta-\alpha;1+\beta;\frac{z}{D}\right)z^{\beta}-z^{\beta}\int_{0}^{\bar{z}/2}\,_{1}F_{1}\!\left(\beta-\alpha;1+\beta;\frac{zt}{D}\right)e^{-t}\,\mbox{d}t.
\]
One can verify that, indeed,
\[
-\left(2\,\frac{\partial^{2}}{\partial z\partial\bar{z}}+\left(1-\frac{z}{D}\right)\frac{\partial}{\partial z}\right)g_{j}\left(z,\bar{z}\right)=\frac{\alpha}{D}\, g_{j}\left(z,\bar{z}\right),\ \text{ }\text{for}\ j=1,2.
\]
For $j=1$ the verification leads to equation (\ref{eq:1F1_ident}).
The case $j=2$ is guaranteed by (\ref{eq:2F1_ident}) and again by
(\ref{eq:1F1_ident}).

We conclude that $\tilde{\psi}_{2}=\psi_{1}+\varphi$ is an exact
formal eigenfunction solving the differential equation $H_{B}^{\text{max}}\tilde{\psi}_{2}=E_{1}\tilde{\psi}_{2}$
on $\mathbb{R}^{2}\backslash L_{a}\cup L_{b}$. Moreover, $\tilde{\psi}_{2}$
fulfills the desired boundary condition on the cut $L_{b}$ exactly
but on $L_{a}$ only approximately. Nevertheless the relative error
of the boundary condition on $L_{a}$ is of order (at most) $D^{\alpha-\beta}e^{-D/2}$.
In fact, referring to (\ref{eq:psi1sol}) (where $\pmb{a}=(0,0)$),
$\psi_{1}$ fulfills the boundary condition on $L_{a}$ exactly. As
far as $\varphi$ is concerned, its behavior for $\left|\theta_{b}\right|<\pi$
is perhaps best seen from the form (\ref{eq:W}).

\section{The shift of the energy \label{sec:DeltaE}}

\subsection{A general formula for $\Delta E$}

Denote
\[
\nabla_{\! A}\!\Big(\frac{\partial}{\partial\mathbf{n}}\Big)=\frac{\partial}{\partial\mathbf{n}}-\frac{ie}{\hbar c}\,\mathbf{n}\cdot\mathbf{A}
\]
where $\mathbf{n}$ is the outer normalized normal vector on the boundary
of a region $G$. Note that Green's second identity is also applicable
to magnetic Laplacians\\
$\Delta_{A}=(\nabla-ie/(\hbar c)\mathbf{A})^{2}$,
\begin{eqnarray*}
\int_{G}\left(\,\overline{f}\,\Delta_{A}g-g\,\overline{\Delta_{A}f}\,\right)\!\mbox{d}S & = & \int_{G}\nabla\!\left(\,\overline{f}\left(\nabla-\frac{ie}{\hbar c}\,\mathbf{A}\right)g-g\,\overline{\left(\nabla-\frac{ie}{\hbar c}\,\mathbf{A}\right)f}\,\right)\!\mbox{d}S=\\
 & = & \int_{\partial G}\left(\,\overline{f}\!\left(\frac{\partial}{\partial\mathbf{n}}-\frac{ie}{\hbar c}\,\mathbf{n}\cdot\mathbf{A}\right)\! g-g\,\overline{\left(\frac{\partial}{\partial\mathbf{n}}-\frac{ie}{\hbar c}\,\mathbf{n}\cdot\mathbf{A}\right)\! f}\,\right)\!\mbox{d}\ell\\
 & = & \int_{\partial G}\left(\,\overline{f}\,\nabla_{\! A}\!\Big(\frac{\partial}{\partial\mathbf{n}}\Big)g-g\,\overline{\nabla_{\! A}\!\Big(\frac{\partial}{\partial\mathbf{n}}\Big)f}\,\right)\!\mbox{d}\ell.
\end{eqnarray*}

Still assuming (\ref{eq:0_alpha_beta_1}). The reader is reminded
that we denote by $\psi_{2}$ an exact two-solenoid eigenfunction
and by $\tilde{\psi}_{2}$ its approximation, cf. (\ref{eq:psi2_W})
and (\ref{eq:W}) or  (\ref{eq:phi_U2}) where $\psi_{1}$ is the
one-solenoid normalized eigenfunction, as given in (\ref{eq:psi1sol})
(with $\pmb{a}=\pmb{0}$). The corresponding exact eigenvalue $E_{2}$
is supposed to be written in the form
\begin{equation}
E_{2}=E_{1}+\Delta E\label{eq:E2_perturb}
\end{equation}
where $E_{1}$, as introduced in (\ref{eq:E1}), is the exact one-solenoid
energy corresponding to $\psi_{1}$.

We shall again work with the maximal differential operator $H_{B}^{\text{max}}$
on the region $\mathbb{R}^{2}\setminus L_{a}\cup L_{b}$ which we
have introduced in Subsection~\ref{subsec:approx_exact_ef}. Then
$\psi_{1}$ and $\psi_{2}$ belong both to the domain of $H_{B}^{\text{max}}$
and both of them are its eigenfunctions with the eigenvalues $E_{1}$
and $E_{2}$, respectively. One observes that the one-solenoid and
the two-solenoid Hamiltonians are both restrictions of $H_{B}^{\text{max}}$.
In the latter case, the domain is determined by the boundary conditions
on $L_{a}\cup L_{b}$, see (\ref{eq:bc}), while in the former case,
the domain is determined by the boundary conditions on $L_{a}$ only
and with no discontinuity being supposed on $L_{b}$ (which corresponds
to letting $\beta=0$ in (\ref{eq:bc})).

Thus we have $H_{B}^{\text{max}}\psi_{1}=E_{1}\psi_{1}$, and $\psi_{1}$
satisfies that part of the boundary conditions which has been imposed
on $L_{a}$ but has no discontinuity on $L_{b}$. Furthermore, $H_{B}^{\text{max}}\psi_{2}=E_{2}\psi_{2}$,
and $\psi_{2}$ satisfies the boundary conditions (\ref{eq:bc}) both
on $L_{a}$ and $L_{b}$. As suggested in (\ref{eq:E2_perturb}),
$E_{2}$ is regarded as a perturbation of $E_{1}$. Let us write
\begin{equation}
\psi_{2}=\tilde{\psi}_{2}+\eta=\psi_{1}+\varphi+\eta.\label{eq:perturb2_eta}
\end{equation}
We have verified, in Subsection~\ref{subsec:approx_exact_ef}, that
the approximate two-solenoid eigenfunction $\tilde{\psi}_{2}$ fulfills
the formal eigenvalue equation $H_{B}^{\text{max}}\tilde{\psi}_{2}=E_{1}\tilde{\psi}_{2}$.
Concerning the boundary conditions, $\tilde{\psi}_{2}$ satisfies
the boundary conditions on $L_{b}$ exactly, see (\ref{eq:psi2t_Lb}),
and on $L_{a}$ only approximately.

To derive a formula for the energy shift $\Delta E$ occurring when
the second solenoid is switched on we suppose that the correction
$\eta$ in (\ref{eq:perturb2_eta}) is comparatively small and, in
particular, that $\langle\psi_{2},\eta\rangle$ can be neglected with
respect to $1$. Starting from the eigenvalue equation
\[
H_{B}^{\text{max}}(\tilde{\psi}_{2}+\eta)=H_{B}^{\text{max}}\psi_{2}=E_{2}\psi_{2}=(E_{1}+\Delta E)(\tilde{\psi}_{2}+\eta),
\]
which simplifies to $H_{B}^{\text{max}}\eta=\Delta E\,\tilde{\psi}_{2}+E_{2}\eta$,
and taking the scalar product with $\psi_{2}$ we obtain
\[
\Delta E\approx\Delta E\,\langle\psi_{2},\tilde{\psi}_{2}\rangle=\langle\psi_{2},H_{B}^{\text{max}}\eta\rangle-\langle H_{B}^{\text{max}}\psi_{2},\eta\rangle.
\]
Since $\psi_{2}$ and $\tilde{\psi}_{2}$ satisfy both the boundary
conditions on $L_{b}$ it is seen from (\ref{eq:perturb2_eta}) that
the same is true for $\eta$. Green's second identity then tells us
that
\begin{equation}
\Delta E\approx\frac{\hbar^{2}}{2\mu}\int_{L_{a}}\left(\eta\,\overline{\nabla_{\! A}\!\left(\frac{\partial}{\partial\mathbf{n}}\right)\!\psi_{2}}-\overline{\psi_{2}}\,\nabla_{\! A}\!\left(\frac{\partial}{\partial\mathbf{n}}\right)\!\eta\right)\mbox{d}\ell.\label{eq:DeltaE_eta}
\end{equation}

Note that the cut $L_{a}$ has two sides, denoted $L_{a}(+)$ and
$L_{a}(-)$, which are determined by the equations $\theta_{a}=\pi$
and $\theta_{a}=-\pi$, respectively. Using this notation we can rewrite
the boundary condition on $L_{a}$ as
\begin{equation}
\psi_{2}\Big|_{L_{a}(+)}=e^{2\pi i\alpha}\,\psi_{2}\Big|_{L_{a}(-)},\ \frac{\partial\psi_{2}}{\partial\mathbf{n}}\Big|_{L_{a}(+)}=-e^{2\pi i\alpha}\,\frac{\partial\psi_{2}}{\partial\mathbf{n}}\Big|_{L_{a}(-)}.\label{eq:bc_La}
\end{equation}
The sign in the latter equation comes from the fact that the outward
oriented normalized normal vectors have opposite signs on $L_{a}(+)$
and $L_{a}(-)$. Consequently, the RHS of (\ref{eq:DeltaE_eta}) equals
\begin{eqnarray}
 &  & \hskip-1em\frac{\hbar^{2}}{2\mu}\int_{L_{a}(+)}\Bigg(\!\left(\eta\bigg|_{L_{a}(+)}-e^{2\pi i\alpha}\,\eta\bigg|_{L_{a}(-)}\right)\!\overline{\nabla_{\! A}\!\left(\frac{\partial}{\partial\mathbf{n}}\right)\!\psi_{2}}\bigg|_{L_{a}(+)}\nonumber \\
 &  & \qquad\qquad-\left(\nabla_{\! A}\!\left(\frac{\partial}{\partial\mathbf{n}}\right)\!\eta\bigg|_{L_{a}(+)}+e^{2\pi i\alpha}\,\nabla_{\! A}\!\left(\frac{\partial}{\partial\mathbf{n}}\right)\!\eta\bigg|_{L_{a}(-)}\right)\overline{\psi_{2}}\bigg|_{L_{a}(+)}\,\Bigg)\mbox{d}\ell.\label{eq:DelataE_bc}
\end{eqnarray}
Recalling that $\psi_{1}$ obeys (\ref{eq:bc_La}), too, and referring
to (\ref{eq:perturb2_eta}) we find that $\eta$ in (\ref{eq:DelataE_bc})
can be replaced by $-\varphi$. Furthermore, still regarding $\eta$
as a small correction to the approximate eigenstate, in (\ref{eq:DelataE_bc})
we replace $\psi_{2}$ by $\tilde{\psi}_{2}=\psi_{1}+\varphi$. But
keeping only the leading asymptotic term as $D$ tends to the infinity
(see (\ref{eq:D})), $\tilde{\psi}_{2}$ can be further reduced to
$\psi_{1}$ (note that $r_{b}>R$ and hence $\varphi$ is, at least,
of order $e^{-D/2}$ on $L_{a}$). Since $\varphi$ has no discontinuity
on $L_{a}$ we finally get
\begin{equation}
\Delta E\approx\frac{\hbar^{2}}{2\mu}\,(1-e^{2\pi i\alpha})\int_{0}^{\infty}\left(-\varphi\,\overline{\nabla_{\! A}\!\left(\frac{\partial}{\partial\mathbf{n}}\right)\!\psi_{1}}\bigg|_{\theta_{a}=\pi}+\overline{\psi_{1}}\bigg|_{\theta_{a}=\pi}\,\nabla_{\! A}\!\left(\frac{\partial}{\partial\mathbf{n}}\right)\!\varphi\right)\!\mbox{d}r_{a}\label{eq:DeltaE_gen}
\end{equation}
where $\mathbf{n}$ is the outward oriented normalized normal vector
on $L_{a}(+)$.

\subsection{Derivation of the formula for the energy shift}

Let us start from (\ref{eq:W}). Recalling (\ref{eq:D}) and again
letting $z=(r_{b}/R)\, e^{i\theta_{b}}$, we can rewrite the integral
in (\ref{eq:W}) for $|\theta_{b}|<\pi/2$ as follows
\begin{eqnarray*}
 &  & \int_{-\infty}^{\infty}(1+ze^{-u})^{\alpha}\exp\!\left(-\frac{D\overline{z}}{2}\, e^{u}\right)\!\frac{e^{\beta u}}{1+e^{u}}\,\mbox{d}u\\
 &  & =\left(\frac{D\overline{z}}{2}\right)^{\!\alpha-\beta}z^{\alpha}\int_{0}^{\infty}\left(1+\frac{2t}{D|z|^{2}}\right)^{\!\alpha}e^{-t}\left(1+\frac{2t}{D\overline{z}}\right)^{\!-1}\, t^{-1+\beta-\alpha}\,\mbox{d}t\\
 &  & =\left(\frac{D}{2}\right)^{\!-\beta/2}\Gamma(\beta-\alpha)\left(\sqrt{\frac{D}{2}}\,\frac{r_{b}}{R}\right)^{\!2\alpha-\beta}e^{i\beta\theta_{b}}\left(1+O(D^{-1})\right)\!.
\end{eqnarray*}
Thus, in the vicinity of the cut $L_{a}$, we have the approximation
\begin{eqnarray}
 &  & \hskip-1em\varphi(\pmb{x})\,=\,-\,\sqrt{\frac{D}{2\pi\Gamma(\alpha+1)}}\frac{\sin(\pi\beta)}{\pi R}\,\Gamma(\beta-\alpha)\!\left(\frac{D}{2}\right)^{\!(\alpha-\beta)/2}\nonumber \\
 &  & \qquad\quad\times\exp\!\left(-\frac{D}{4R^{2}}\left(R^{2}+r_{b}^{2}\right)+\frac{iD}{2R}\, r_{b}\sin\left(\theta_{b}\right)\!\right)\!\left(\sqrt{\frac{D}{2}}\,\frac{r_{b}}{R}\right)^{\!2\alpha-\beta}e^{i\beta\theta_{b}}\left(1+O(D^{-1})\right)\!.\nonumber \\
\label{eq:phi_La}
\end{eqnarray}

On $L_{a}(+)$, $\theta_{a}=\pi$, $\theta_{b}=0$ and
\[
\frac{\partial}{\partial\mathbf{n}}=-\frac{\partial}{\partial x_{2}}=\frac{1}{r_{a}}\,\frac{\partial}{\partial\theta_{a}}=\frac{1}{r_{b}}\,\frac{\partial}{\partial\theta_{b}}\,.
\]
At the same time, referring to (\ref{eq:avect_uni}), $\pmb{A}_{B}=(B/2)\, r_{a}^{\,2}\,\nabla\theta_{a}$
whence
\[
\nabla_{\! A}\!\left(\frac{\partial}{\partial\mathbf{n}}\right)=\frac{1}{r_{a}}\,\frac{\partial}{\partial\theta_{a}}-\frac{ieB}{2\hbar c}\, r_{a}.
\]
In view of (\ref{eq:polar_ab}), approximation (\ref{eq:phi_La})
can be rewritten in the form
\begin{eqnarray*}
\varphi(\pmb{x}) & \approx & -\,\sqrt{\frac{D}{2\pi\Gamma(\alpha+1)}}\frac{\sin(\pi\beta)}{\pi R}\,\Gamma(\beta-\alpha)\!\left(\frac{D}{2}\right)^{\!(\alpha-\beta)/2}\\
 &  & \times\exp\!\left(-\frac{D}{2}-\frac{Dr_{a}^{\,2}}{4R^{2}}+\frac{Dr_{a}}{2R}\, e^{-i\theta_{a}}\right)\!\left(\sqrt{\frac{D}{2}}\,\frac{r_{b}}{R}\right)^{\!2\alpha-\beta}(R-r_{a}e^{i\theta_{a}})^{\beta}\, r_{b}^{\,-\beta}.
\end{eqnarray*}
It follows that
\[
\nabla_{\! A}\!\left(\frac{\partial}{\partial\mathbf{n}}\right)\!\varphi(\pmb{x})\approx\varphi(\pmb{x})\!\left(\frac{iD}{2R}+\frac{i\beta}{r_{b}}+\frac{iDr_{a}}{2R^{2}}\right)=\varphi(\pmb{x})\,\frac{iDr_{b}}{2R^{2}}\left(1+O(D^{-1})\right)\!.
\]
Recalling (\ref{eq:psi1sol}),
\[
\nabla_{\! A}\!\left(\frac{\partial}{\partial\mathbf{n}}\right)\!\psi_{1}(\pmb{x})=\psi_{1}(\pmb{x})\!\left(\frac{i\alpha}{r_{a}}+\frac{iDr_{a}}{2R^{2}}\right)\!.
\]
Plugging these relations into (\ref{eq:DeltaE_gen}) we obtain
\begin{eqnarray*}
\Delta E & \approx & \frac{i\hbar^{2}}{2\mu R}\,\sqrt{\frac{D}{2}}\,(e^{-i\pi\alpha}-e^{i\pi\alpha})\,\frac{(-1)\mu\omega_{c}}{2\pi\Gamma(\alpha+1)\hbar}\frac{\sin(\pi\beta)}{\pi}\,\Gamma(\beta-\alpha)\!\left(\frac{D}{2}\right)^{\!(\alpha-\beta)/2}e^{-D/2}\\
 &  & \times\int_{0}^{\infty}\exp\!\left(-\frac{Dr_{a}^{\,2}}{2R^{2}}-\frac{Dr_{a}}{2R}\right)\!\left(\sqrt{\frac{D}{2}}\,\frac{r_{b}}{R}\right)^{\!2\alpha-\beta}\\
 &  & \qquad\times\left(\sqrt{\frac{D}{2}}\,\frac{r_{a}}{R}+\alpha\!\left(\sqrt{\frac{D}{2}}\,\frac{r_{a}}{R}\right)^{\!-1}+\sqrt{\frac{D}{2}}\,\frac{r_{b}}{R}\right)\!\left(\sqrt{\frac{D}{2}}\,\frac{r_{a}}{R}\right)^{\!\alpha}\mbox{d}r_{a}.
\end{eqnarray*}
Applying the substitution $r_{a}=2Rx/D$ we arrive at the expression
\begin{eqnarray*}
\Delta E & \approx & -\frac{\hbar\omega_{c}\sin(\pi\alpha)\sin(\pi\beta)}{2\pi^{2}\,\Gamma(\alpha+1)}\,\Gamma(\beta-\alpha)\!\left(\frac{D}{2}\right)^{\!\alpha-\beta}e^{-D/2}\int_{0}^{\infty}e^{-x}\left(\frac{\alpha}{x}+1\right)x^{\alpha}\,\mbox{d}x\\
 &  & \times\left(1+O(D^{-1})\right).
\end{eqnarray*}

We conclude that the desired formula for the energy shift reads (see
(\ref{eq:D}))
\[
\Delta E\approx-\frac{\sin(\pi\alpha)\sin(\pi\beta)}{\pi^{2}}\,\Gamma(\beta-\alpha)\!\left(\frac{D}{2}\right)^{\!\alpha-\beta}e^{-D/2}\,\hbar\omega_{c}.
\]

\section*{Acknowledgments}

The author wishes to acknowledge gratefully partial support from grant
No. GA13-11058S of the Czech Science Foundation.

\setcounter{section}{1}
\renewcommand{\thesection}{\Alph{section}}
\setcounter{equation}{0}
\renewcommand{\theequation}{\Alph{section}.\arabic{equation}}

\section*{Appendix~\thesection. Hypergeometric functions, the incomplete
gamma function}

Here we collect, for the reader's convenience, several formulas concerned
with some special functions playing an important role in the derivations
throughout the paper.

Euler's transformation formulas for the hypergeometric function tell
us that \cite[Eq.~9.131(1)]{GradshteynRyzhik}:
\begin{equation}
\,_{2}F_{1}(a,b;c;z)=(1-z)^{-a}\,\,_{2}F_{1}\!\left(a,c-b;c;\frac{z}{z-1}\right)\label{eq:Euler1}
\end{equation}
and
\begin{equation}
\,_{2}F_{1}(a,b;c;z)=(1-z)^{c-a-b}\,\,_{2}F_{1}(c-a,c-b;c;z).\label{eq:Euler2}
\end{equation}
By another transformation formula \cite[Eq. 9.131(2)]{GradshteynRyzhik},
\begin{eqnarray}
\,_{2}F_{1}(a,b;c;1-z) & = & \frac{\Gamma(c)\Gamma(c-a-b)}{\Gamma(c-a)\Gamma(c-b)}\,\,_{2}F_{1}(a,b;a+b-c+1;z)\nonumber \\
 &  & +\, z^{c-a-b}\,\frac{\Gamma(c)\Gamma(a+b-c)}{\Gamma(a)\Gamma(b)}\,\,_{2}F_{1}(c-a,c-b;c-a-b+1;z),\nonumber \\
\noalign{\vskip-5pt}\label{eq:hyperfce_transf}
\end{eqnarray}

Let us recall, too, the following two basic identities for the hypergeometric
function and the confluent hypergeometric function,
\begin{eqnarray}
 &  & -\, b\,_{2}F_{1}(1,1+b-a;2+b;z)-(1-z)\,\,_{2}F_{1}(2,1+b-a;2+b;z)\nonumber \\
 &  & +\,(1+b)\,\,_{2}F_{1}(1,b-a;1+b;z)\,=\,0\label{eq:2F1_ident}
\end{eqnarray}
and
\begin{equation}
\,_{1}F_{1}(a;c;x)-\left(1-\frac{x}{c}\right)\,_{1}F_{1}(a+1;c+1;x)-\frac{(a+1)x}{c\,(c+1)}\,\,_{1}F_{1}(a+2;c+2;x)=0.\label{eq:1F1_ident}
\end{equation}

The second confluent hypergeometric function $U(a,b;z)$ (in the literature
alternatively denoted $\Psi(a,b;c)$), fulfills \cite[Eq. 9.210(2)]{GradshteynRyzhik}
\begin{eqnarray}
U(a,c,z) & = & \frac{\pi}{\sin(\pi c)}\left(\frac{\,_{1}F_{1}(a;c;z)}{\Gamma(c)\Gamma(1+a-c)}-\frac{z^{1-c}\,_{1}F_{1}(1+a-c;2-c;z)}{\Gamma(a)\Gamma(2-c)}\right)\nonumber \\
\noalign{\smallskip} & = & z^{1-c}\, U(1+a-c;2-c;z).\label{eq:U}
\end{eqnarray}

Let us further remark that
\[
\sum_{m=0}^{\infty}\frac{r^{m+\sigma}}{\Gamma(m+\sigma+1)}=\frac{1}{\Gamma(\sigma)}\, e^{r}\int_{0}^{r}t^{-1+\sigma}e^{-t}\,\mbox{d}t=e^{r}-\frac{e^{r}\,\Gamma(\sigma,r)}{\Gamma(\sigma)}
\]
holds for $\sigma>0$ and $r\geq0$ where
\[
\Gamma(\sigma,z)=\int_{z}^{\infty}t^{\sigma-1}e^{-t}\,\mbox{d}t
\]
is the incomplete gamma function. Moreover,
\begin{eqnarray}
\Gamma(\sigma,r) & = & \Gamma(\sigma)\!\left(1-e^{-r}\sum_{m=0}^{\infty}\frac{r^{m+\sigma}}{\Gamma(m+\sigma+1)}\right)=\Gamma(\sigma)-\sum_{k=0}^{\infty}\frac{(-1)^{k}r^{k+\sigma}}{k!\,(k+\sigma)}\nonumber \\
 & = & \Gamma(\sigma)-\frac{1}{\sigma}\,\,_{1}F_{1}(\sigma;1+\sigma;-r)\, r^{\sigma}\label{eq:Gamma_incompl3}
\end{eqnarray}
holds for all $r>0$ and $\sigma\in\mathbb{C}$, $\sigma\neq0,-1,-2,\ldots$.

\addtocounter{section}{1}
\manuallabel{app:asympt}{\thesection}
\setcounter{equation}{0}
\setcounter{step}{0}
\renewcommand{\thestep}{\arabic{step}}

\section*{Appendix~\thesection. An asymptotic formula}

Suppose $X,Y,Z>0$, $0<\alpha<\beta<1$ and $\phi_{1},\phi_{2}\in(-\pi,\pi)$.
Let
\begin{eqnarray*}
V(\epsilon,u_{1},u_{2}) & = & \exp\!\big(-(Xe^{u_{1}}+Ye^{u_{2}}+Ze^{u_{1}+u_{2}})\\
 &  & \qquad-2\epsilon\left(X\cosh(u_{1})+Y\cosh(u_{2})+Z\cosh(u_{1}+u_{2})\right)\big)\\
 &  & \qquad\qquad\qquad\times\,\frac{e^{\alpha u_{1}+\beta u_{2}}}{(1+e^{u_{1}+i\phi_{1}})(1+e^{u_{2}+i\phi_{2}})}\,.
\end{eqnarray*}
The dependence of $V$ on $X$, $Y$, $Z$, $\alpha$, $\beta$ and
$\phi$ is not explicitly indicated in the notation.

\begin{proposition*}\label{lemma:asympt_double} Under the above
assumptions,
\begin{eqnarray}
\int_{-\infty}^{\infty}\int_{-\infty}^{\infty}V(\epsilon,u_{1},u_{2})\,\mbox{d}u_{1}\mbox{d}u_{2} & = & \int_{-\infty}^{\infty}\int_{-\infty}^{\infty}V(0,u_{1},u_{2})\,\mbox{d}u_{1}\mbox{d}u_{2}\nonumber \\
\noalign{\smallskip} &  & \hskip-0.2em+\,\Gamma(-\alpha)\epsilon^{\alpha}\int_{-\infty}^{\infty}(X+Ze^{-u})^{\alpha}\exp(-Ye^{u})\,\frac{e^{\beta u}}{1+e^{u+i\phi}}\,\mbox{d}u\nonumber \\
 &  & +\, O(\epsilon^{\beta})\ \ \text{{as}}\ \epsilon\to0+.\label{eq:intV_asympt_tot}
\end{eqnarray}
\end{proposition*}

A proof of the proposition is provided in a sketchy form since it
is somewhat lengthy and tedious though based on routine reasoning.

\begin{step}\label{lemma:asympt_int0} Suppose $0<\alpha<1$ and
$F(u)=e^{-\alpha u}\left(1+O(e^{-u})\right)$. Then
\[
\int_{0}^{\infty}\exp(-2\epsilon\cosh(u))F(u)\,\mbox{d}u=\int_{0}^{\infty}F(u)\,\mbox{d}u+\Gamma(-\alpha)\,\epsilon^{\alpha}+O(\epsilon).
\]
In particular, for $X>0$, $0<\alpha<1$ and $\phi\in(-\pi,\pi)$,
\begin{eqnarray}
 &  & \int_{-\infty}^{\infty}\exp\!\big(-Xe^{u}-2\epsilon X\cosh(u)\big)\frac{e^{\alpha u}}{1+e^{u+i\phi}}\,\mbox{d}u\nonumber \\
 &  & =\,\int_{-\infty}^{\infty}\exp(-Xe^{u})\,\frac{e^{\alpha u}}{1+e^{u+i\phi}}\,\mbox{d}u+\Gamma(-\alpha)\, X^{\alpha}\epsilon^{\alpha}+O(\epsilon).\label{eq:asympt_single}
\end{eqnarray}
\end{step}

In fact, the integral asymptotically equals
\[
\int_{0}^{\infty}\exp(-\epsilon e^{u})F(u)\,\mbox{d}u+O(\epsilon)=\int_{0}^{\infty}\exp(-\epsilon e^{u})e^{-\alpha u}\,\mbox{d}u+\int_{0}^{\infty}\!\big(F(u)-e^{-\alpha u}\big)\mbox{d}u+O(\epsilon),
\]
and from (\ref{eq:Gamma_incompl3}) one deduces that
\[
\int_{0}^{\infty}\exp(-\epsilon e^{u})e^{-\alpha u}\,\mbox{d}u=\epsilon^{\alpha}\Gamma(-\alpha,\epsilon)=\frac{1}{\alpha}+\Gamma(-\alpha)\,\epsilon^{\alpha}+O(\epsilon).
\]

\begin{step}\label{lemma:asympt_int1} Suppose $\epsilon,\sigma,\nu>0$
and $a,b>0$. Then
\begin{eqnarray}
\int_{a}^{\infty}\int_{b}^{\infty}e^{-\epsilon t_{1}t_{2}}t_{1}{}^{-1-\sigma}t_{2}{}^{-1-\nu}\,\mbox{d}t_{1}\mbox{d}t_{2} & = & \frac{1}{\sigma-\nu}\left(a^{-\sigma+\nu}\,\Gamma(-\nu)\,\epsilon^{\nu}-b^{\sigma-\nu}\,\Gamma(-\sigma)\,\epsilon^{\sigma}\right)\nonumber \\
 &  & +\, a^{-\sigma}b^{-\nu}\sum_{k=0}^{\infty}\frac{(-1)^{k}(ab\epsilon)^{k}}{k!(k-\sigma)(k-\nu)}\label{eq:step2}
\end{eqnarray}
(the singularities on the RHS occurring for $\sigma=\nu$ or $\sigma\in\mathbb{N}$
or $\nu\in\mathbb{N}$ can be treated in the limit). \end{step}

A simple scaling of the integrand shows that it is sufficient to consider
the particular case with $a=b=1$. If so, after application of the
transformation of variables $t_{1}=\sqrt{z_{1}z_{2}}$, $t_{2}=\sqrt{z_{1}/z_{2}}$,
one obtains the expression
\[
\frac{1}{2}\int_{1}^{\infty}e^{-\epsilon z_{1}}\, z_{1}^{\,-1-(\sigma+\nu)/2}\left(\int_{1/z_{1}}^{z_{1}}z_{2}^{\,-1-(\sigma-\nu)/2}\,\mbox{d}z_{2}\right)\!\mbox{d}z_{1}=\int_{1}^{\infty}e^{-\epsilon z_{1}}\!\left(z_{1}^{\,-1-\nu}-z_{1}^{\,-1-\sigma}\right)\mbox{d}z_{1}.
\]
Now it is sufficient to apply (\ref{eq:Gamma_incompl3}).

\begin{step}\label{lemma:asympt_int2} Suppose $X$, $Y$, $Z$ are
positive and $0<\alpha,\beta<1$. Then
\begin{eqnarray}
 &  & \int_{0}^{\infty}\int_{0}^{\infty}\exp\!\big(-\epsilon\,(Xe^{u_{1}}+Ye^{u_{2}}+Ze^{u_{1}+u_{2}})\big)e^{-\alpha u_{1}-\beta u_{2}}\,\mbox{d}u_{1}\mbox{d}u_{2}\nonumber \\
 &  & =\,\frac{1}{\alpha\beta}+\frac{\Gamma(-\alpha)(X+Z)^{\alpha}}{\beta-\alpha}\,\,_{2}F_{1}\!\left(1,-\alpha;1-\alpha+\beta;\frac{X}{X+Z}\right)\epsilon^{\alpha}\label{eq:step3}\\
 &  & \quad+\,\frac{\Gamma(-\beta)(Y+Z)^{\beta}}{\alpha-\beta}\,\,_{2}F_{1}\!\left(1,-\beta;1+\alpha-\beta;\frac{Y}{Y+Z}\right)\epsilon^{\beta}+O(\epsilon).\nonumber 
\end{eqnarray}
\end{step}

First rewrite the integral as
\begin{eqnarray*}
 &  & e^{\epsilon XY/Z}\int_{1+Y/Z}^{\infty}\int_{1+X/Z}^{\infty}\exp\!\big(-\epsilon Z\, t_{1}t_{2}\big)\left(t_{1}-\frac{Y}{Z}\right)^{\!-1-\alpha}\!\left(t_{2}-\frac{X}{Z}\right)^{\!-1-\beta}\mbox{d}t_{1}\mbox{d}t_{2}\\
 &  & =\, e^{\epsilon XY/Z}\sum_{k=0}^{\infty}\sum_{\ell=0}^{\infty}\binom{-1-\alpha}{k}\binom{-1-\beta}{\ell}\left(-\frac{Y}{Z}\right)^{\! k}\!\left(-\frac{X}{Z}\right)^{\!\ell}\\
 &  & \qquad\qquad\qquad\times\,\int_{1+Y/Z}^{\infty}\int_{1+X/Z}^{\infty}e^{-\epsilon Z\, t_{1}t_{2}}\, t_{1}^{-1-k-\alpha}\, t_{2}^{-1-\ell-\beta}\,\mbox{d}t_{1}\mbox{d}t_{2}.
\end{eqnarray*}
Referring to (\ref{eq:step2}), we have the asymptotic behavior of
the last integral
\begin{eqnarray*}
 &  & \frac{1}{\alpha\beta}+\sum_{k=0}^{\infty}\binom{-1-\alpha}{k}\left(-\frac{Y}{Z}\right)^{\! k}\frac{1}{k+\alpha-\beta}\left(\frac{Z}{Y+Z}\right)^{\! k+\alpha-\beta}\Gamma(-\beta)(Z\epsilon)^{\beta}\\
 &  & +\sum_{\ell=0}^{\infty}\binom{-1-\beta}{\ell}\left(-\frac{X}{Z}\right)^{\!\ell}\frac{1}{\ell-\alpha+\beta}\left(\frac{Z}{X+Z}\right)^{\!\ell+\alpha-\beta}\Gamma(-\alpha)(Z\epsilon)^{\alpha}+O(\epsilon).
\end{eqnarray*}
To complete the derivation it suffices to apply the very definition
of the hypergeometric function and (\ref{eq:Euler2}).

\begin{step}\label{lemma:asympt_int4} Suppose $0<\gamma<1$, $\sigma>2$,
$a,b>0$, and $F(t)=O(t^{-\sigma})$ as $t\to+\infty$. Then
\begin{equation}
\int_{a}^{\infty}\int_{b}^{\infty}e^{-\epsilon t_{1}t_{2}}F(t_{2})t_{1}^{\,-1-\gamma}\,\mbox{d}t_{1}\mbox{d}t_{2}=\frac{a^{-\gamma}}{\gamma}\int_{b}^{\infty}F(t)\,\mbox{d}t+\Gamma(-\gamma)\!\left(\int_{b}^{\infty}F(t)t^{\gamma}\,\mbox{d}t\right)\!\epsilon^{\gamma}+O(\epsilon)\label{eq:step4}
\end{equation}
\end{step}

A simple scaling of the integrand shows that it suffices to consider
the particular case $a=b=1$. Then the LHS equals
\begin{eqnarray*}
 &  & \hskip-2em\epsilon^{\gamma}\int_{1}^{\infty}F(t)t^{\gamma}\!\left(\int_{\epsilon t}^{\infty}e^{-u}\, u^{-1-\gamma}\,\mbox{d}u\right)\!\mbox{d}t\\
 &  & \hskip7em=\,\int_{1}^{\infty}F(t)\!\left(\frac{1}{\gamma}\, e^{-\epsilon t}-\frac{(\epsilon t)^{\gamma}}{\gamma}\,\Gamma(1-\gamma)+\frac{(\epsilon t)^{\gamma}}{\gamma}\int_{0}^{\epsilon t}e^{-u}\, u^{-\gamma}\,\mbox{d}u\right)\!\mbox{d}t.
\end{eqnarray*}
Now it suffices to observe that
\[
(\epsilon t)^{\gamma}\int_{0}^{\epsilon t}e^{-u}\, u^{-\gamma}\,\mbox{d}u\leq\frac{\epsilon t}{1-\gamma}\,.
\]

\begin{step}\label{lemma:asympt_int3} Suppose $X,Z>0$, $0<\alpha<1<\sigma$,
and $F(u)=O(e^{-\sigma u})$ as $u\to+\infty$. Then
\begin{eqnarray}
 &  & \int_{0}^{\infty}\int_{0}^{\infty}\exp\!\big(-\epsilon\,(Xe^{u_{1}}+Ze^{u_{1}+u_{2}})\big)F(u_{2})\, e^{-\alpha u_{1}}\,\mbox{d}u_{1}\mbox{d}u_{2}\nonumber \\
 &  & =\,\frac{1}{\alpha}\int_{0}^{\infty}F(u)\,\mbox{d}u+\left(\int_{0}^{\infty}\left(Ze^{u}+X\right)^{\alpha}F(u)\, du\right)\!\Gamma(-\alpha)\,\epsilon^{\alpha}+O(\epsilon).\label{eq:step5}
\end{eqnarray}
\end{step}

Let $G(t)=F(\ln t)/t=O(t^{-1-\sigma})$. The integral equals
\[
\int_{1}^{\infty}\int_{1+X/Z}^{\infty}\exp(-\epsilon Zt_{1}t_{2})\, G\!\left(t_{2}-\frac{X}{Z}\right)\! t_{1}^{\,-1-\alpha}\,\mbox{d}t_{1}\mbox{d}t_{2},
\]
and according to (\ref{eq:step4}), this expression is equal to
\[
\frac{1}{\alpha}\int_{1+X/Z}^{\infty}G\!\left(t-\frac{X}{Z}\right)\!\mbox{d}t+\Gamma(-\alpha)\!\left(\int_{1+X/Z}^{\infty}G\!\left(t-\frac{X}{Z}\right)t^{\alpha}\,\mbox{d}t\right)\!(\epsilon Z)^{\alpha}+O(\epsilon).
\]

\begin{step}\label{lemma:asympt_int5} For $X,Z>0$, $c>-1$ and
$d$ arbitrary it holds true that
\begin{equation}
\int_{0}^{1}(Xt+Z)^{d}\, t^{c}\,\mbox{d}t=\frac{(X+Z)^{d}}{1+c}\,\,_{2}F_{1}\!\left(1,-d;c+2;\frac{X}{X+Z}\right)\!,\label{eq:step6}
\end{equation}
see \cite[Eq.~3.194(1)]{GradshteynRyzhik} and (\ref{eq:Euler1}).
\end{step}

\begin{step}\label{lemma:asympt_minusminus} Assume that $X,Y,Z>0$,
$0<\alpha<\beta<1$ and $\phi_{1},\phi_{2}\in(-\pi,\pi)$. Then
\begin{eqnarray}
 &  & \hskip-2em\int_{0}^{\infty}\int_{0}^{\infty}V(\epsilon,-u_{1},-u_{2})\,\mbox{d}u_{1}\mbox{d}u_{2}\,=\,\int_{0}^{\infty}\int_{0}^{\infty}V(0,-u_{1},-u_{2})\,\mbox{d}u_{1}\mbox{d}u_{2}\nonumber \\
 &  & \hskip7em+\,\Gamma(-\alpha)\!\left(\int_{0}^{\infty}\left(Ze^{u}+X\right)^{\alpha}\frac{\exp(-Ye^{-u})\, e^{-\beta u}}{1+e^{-u+i\phi_{2}}}\,\mbox{d}u\right)\!\epsilon^{\alpha}+O(\epsilon^{\beta}).\nonumber \\
\label{eq:int_Vminusminus}
\end{eqnarray}
\end{step}

For simplicity we put $\phi_{1}=\phi_{2}=0$ but the manipulations
to follow can readily be extended to arbitrary values $\phi_{1},\phi_{2}\in(-\pi,\pi)$.
The integral asymptotically equals
\begin{eqnarray*}
 &  & \hskip-1.3em\int_{0}^{\infty}\int_{0}^{\infty}\!\left(\exp(-Ze^{-u_{1}-u_{2}})-1\right)\exp\!\big(-(Xe^{-u_{1}}+Ye^{-u_{2}})\big)\frac{e^{-\alpha u_{1}-\beta u_{2}}}{(1+e^{-u_{1}})(1+e^{-u_{2}})}\,\mbox{d}u_{1}\mbox{d}u_{2}\\
 &  & \hskip-1.3em+\,\int_{0}^{\infty}\int_{0}^{\infty}\exp\!\big(-\epsilon\,(Xe^{u_{1}}+Ye^{u_{2}}+Ze^{u_{1}+u_{2}})\big)\exp\!\big(-(Xe^{-u_{1}}+Ye^{-u_{2}})\big)\\
 &  & \qquad\quad\times\,\frac{e^{-\alpha u_{1}-\beta u_{2}}}{(1+e^{-u_{1}})(1+e^{-u_{2}})}\,\mbox{d}u_{1}\mbox{d}u_{2}+O(\epsilon).
\end{eqnarray*}
The second integral in this expression equals
\begin{eqnarray*}
 &  & \hskip-1.3em\int_{0}^{\infty}\int_{0}^{\infty}\exp\!\big(-\epsilon\,(Xe^{u_{1}}+Ye^{u_{2}}+Ze^{u_{1}+u_{2}})\big)e^{-\alpha u_{1}-\beta u_{2}}\,\mbox{d}u_{1}\mbox{d}u_{2}\\
 &  & \hskip-1.3em-\int_{0}^{\infty}\int_{0}^{\infty}\exp\!\big(-\epsilon\,(Ye^{u_{2}}+Ze^{u_{1}+u_{2}})\big)\!\left(1-\frac{\exp(-Xe^{-u_{1}})}{1+e^{-u_{1}}}\right)e^{-\alpha u_{1}-\beta u_{2}}\,\mbox{d}u_{1}\mbox{d}u_{2}\\
 &  & \hskip-1.3em-\int_{0}^{\infty}\int_{0}^{\infty}\exp\!\big(-\epsilon\,(Xe^{u_{1}}+Ze^{u_{1}+u_{2}})\big)\!\left(1-\frac{\exp(-Ye^{-u_{2}})}{1+e^{-u_{2}}}\right)e^{-\alpha u_{1}-\beta u_{2}}\,\mbox{d}u_{1}\mbox{d}u_{2}\\
 &  & \hskip-1.3em+\int_{0}^{\infty}\int_{0}^{\infty}\left(1-\frac{\exp(-Xe^{-u_{1}})}{1+e^{-u_{1}}}\right)\!\left(1-\frac{\exp(-Ye^{-u_{2}})}{1+e^{-u_{2}}}\right)e^{-\alpha u_{1}-\beta u_{2}}\,\mbox{d}u_{1}\mbox{d}u_{2}+O(\epsilon).
\end{eqnarray*}
To complete the derivation one can apply (\ref{eq:step3}) to the
first integral and (\ref{eq:step5}) to the second and third integral
in the last expression while taking into account (\ref{eq:step6}).

\begin{step}\label{lemma:asympt_minusplus} Assume $X,Y,Z>0$, $0<\alpha,\beta<1$
and $\phi_{1},\phi_{2}\in(-\pi,\pi)$. Then
\begin{eqnarray}
 &  & \hskip-3em\int_{0}^{\infty}\int_{0}^{\infty}V(\epsilon,-u_{1},u_{2})\,\mbox{d}u_{1}\mbox{d}u_{2}\,=\,\int_{0}^{\infty}\int_{0}^{\infty}V(0,-u_{1},u_{2})\,\mbox{d}u_{1}\mbox{d}u_{2}\nonumber \\
 &  & \qquad\qquad\qquad+\,\Gamma(-\alpha)\!\left(\int_{0}^{\infty}\left(Ze^{-u}+X\right)^{\alpha}\exp(-Ye^{u})\,\frac{e^{\beta u}}{1+e^{u+i\phi_{2}}}\,\mbox{d}u\right)\!\epsilon^{\alpha}+O(\epsilon).\nonumber \\
\label{eq:int_Vminusplus}
\end{eqnarray}
\end{step}

To simplify the notation let us consider just the case $\phi_{1}=\phi_{2}=0$.
The integral asymptotically equals
\begin{eqnarray*}
 &  & \hskip-1.8em\int_{0}^{\infty}\int_{0}^{\infty}\exp\!\big(-(Xe^{-u_{1}}+Ye^{u_{2}}+Ze^{-u_{1}+u_{2}})-\epsilon\,(Xe^{u_{1}}+Ze^{u_{1}-u_{2}})\big)\\
 &  & \hskip-1.8em\qquad\qquad\times\,\frac{e^{-\alpha u_{1}-(1-\beta)u_{2}}}{(1+e^{-u_{1}})(1+e^{-u_{2}})}\,\mbox{d}u_{1}\mbox{d}u_{2}+O(\epsilon)\\
 &  & \hskip-1.8em=\int_{0}^{\infty}\int_{0}^{\infty}\!\left(\frac{\exp(-Xe^{-u_{1}})}{1+e^{-u_{1}}}-1\right)\!\exp\!\big(-(Ye^{u_{2}}+Ze^{-u_{1}+u_{2}})\big)\frac{e^{-\alpha u_{1}-(1-\beta)u_{2}}}{1+e^{-u_{2}}}\,\mbox{d}u_{1}\mbox{d}u_{2}\\
 &  & \hskip-1.8em\quad+\,\int_{0}^{\infty}\int_{0}^{\infty}\exp\!\big(-(Ye^{u_{2}}+Ze^{-u_{1}+u_{2}})-\epsilon\,(Xe^{u_{1}}+Ze^{u_{1}-u_{2}})\big)\frac{e^{-\alpha u_{1}-(1-\beta)u_{2}}}{1+e^{-u_{2}}}\,\mbox{d}u_{1}\mbox{d}u_{2}\\
\noalign{\smallskip} &  & \hskip-1.8em\quad+\, O(\epsilon).
\end{eqnarray*}
In the very last integral in this equation we split the integration
domain into two disjoint subdomains, $u_{1}<u_{2}$ and $u_{1}>u_{2}$,
and apply the substitutions $u_{2}=u_{1}+v_{2}$ and $u_{1}=u_{2}+v_{1}$,
respectively. The derivation then goes on in a very routine manner.

\noindent\addtocounter{step}{1}\textbf{$\textbf{{Step\ \thestep.}}$}
It is obvious that
\begin{equation}
\int_{0}^{\infty}\int_{0}^{\infty}V(\epsilon,u_{1},u_{2})\,\mbox{d}u_{1}\mbox{d}u_{2}=\int_{0}^{\infty}\int_{0}^{\infty}V(0,u_{1},u_{2})\,\mbox{d}u_{1}\mbox{d}u_{2}+O(\epsilon).\label{eq:int_Vplusplus}
\end{equation}
Splitting the integration domain on the LHS of (\ref{eq:intV_asympt_tot}),
i.e. the $u_{1}$,$u_{2}$ plane, into four quadrants and applying
(\ref{eq:int_Vplusplus}), (\ref{eq:int_Vminusminus}) and (\ref{eq:int_Vminusplus})
we obtain the result.

\addtocounter{section}{1}
\manuallabel{app:int}{\thesection}
\setcounter{equation}{0}
\setcounter{step}{0}
\renewcommand{\thestep}{\arabic{step}}

\section*{Appendix~\thesection. Evaluation of an integral}

In this appendix it is shown that
\begin{eqnarray}
 &  & \int_{0}^{\infty}\frac{(t+c)^{\alpha}\, t^{\beta-\alpha-1}}{1+t}\, e^{-zt}\,\mbox{d}t\nonumber \\
 &  & =\frac{\pi}{\sin(\pi\beta)}\!\left(\!(1-c)^{\alpha}-\frac{\Gamma(\beta-\alpha)}{\Gamma(-\alpha)\Gamma(1+\beta)}\,\,_{2}F_{1}(1,\beta-\alpha;1+\beta;c)\, c^{\beta}\right)\! e^{z}\label{eq:int_U}\\
 &  & \quad-\,\Gamma(\beta-\alpha)\, z^{1-\beta}\int_{0}^{1}e^{zt}U\big(-\alpha,1-\beta;cz(1-t)\big)(1-t)^{-\beta}\,\mbox{d}t.\nonumber 
\end{eqnarray}
holds for $\alpha<\beta<1$, $0<|c|<1$, $|\arg c|<\pi$, and $\Re z\geq0$.

First, according to \cite[Eq. 3.227]{GradshteynRyzhik}, if $0<\Re\nu<\Re\sigma$,
$\beta,\gamma\neq0$, $|\arg\beta|<\pi$, $|\arg\gamma|<\pi$, then
\begin{equation}
\int_{0}^{\infty}\frac{x^{\nu-1}(\beta+x)^{1-\sigma}}{\gamma+x}\,\mbox{d}x=\beta^{1-\sigma}\gamma^{\nu-1}B(\nu,\sigma-\nu)\,_{2}F_{1}\!\left(\sigma-1,\nu;\sigma;1-\frac{\gamma}{\beta}\right)\!.\label{eq:Beta_2F1}
\end{equation}
Combining (\ref{eq:Beta_2F1}) with (\ref{eq:Euler1}) and (\ref{eq:hyperfce_transf})
we deduce that
\begin{eqnarray}
 &  & \int_{0}^{\infty}\frac{(t+c)^{\alpha}\, t^{\beta-\alpha-1}}{1+t}\,\mbox{d}t\nonumber \\
 &  & =\frac{\pi}{\sin(\pi\beta)}\left(\!(1-c)^{\alpha}-\frac{\Gamma(\beta-\alpha)}{\Gamma(-\alpha)\Gamma(1+\beta)}\,\,_{2}F_{1}(1,\beta-\alpha;1+\beta;c)\, c^{\beta}\right)\label{eq:int_F0}
\end{eqnarray}
holds for for $0<|c|<1$ and $|\arg c|<\pi$.

Second, suppose $\alpha$, $\beta$, $c$ are fixed and put
\[
F(z)=\int_{0}^{\infty}\frac{(t+c)^{\alpha}\, t^{\beta-\alpha-1}}{1+t}\, e^{-zt}\,\mbox{d}t.
\]
Then $F(z)$ is continuous on the half-plane $\Re z\geq0$, analytic
on $\Re z>0$, and obeys the differential equation
\[
F'(z)-F(z)=-G(z)
\]
where
\[
G(z)=\int_{0}^{\infty}(t+c)^{\alpha}\, t^{\beta-\alpha-1}\, e^{-zt}\,\mbox{d}t
\]
Referring to (\ref{eq:U}) and \cite[Eq. 9.211(4)]{GradshteynRyzhik}
we have
\[
G(z)=\Gamma(\beta-\alpha)c^{\beta}U(\beta-\alpha,1+\beta;cz)=\Gamma(\beta-\alpha)z^{-\beta}U(-\alpha,1-\beta;cz).
\]
It follows that
\[
F(z)=F(0)e^{z}-z\int_{0}^{1}e^{zt}G\big(z(1-t)\big)\,\mbox{d}t.
\]
Now it suffices to observe that, by (\ref{eq:int_F0}),
\[
F(0)=\frac{\pi}{\sin(\pi\beta)}\left(\!(1-c)^{\alpha}-\frac{\Gamma(\beta-\alpha)}{\Gamma(-\alpha)\Gamma(1+\beta)}\,\,_{2}F_{1}(1,\beta-\alpha;1+\beta;c)\, c^{\beta}\right)\!.
\]

\newpage{}

\renewcommand{\thefigure}{\arabic{figure}}

\begin{figure}
\includegraphics[width=0.95\textwidth]{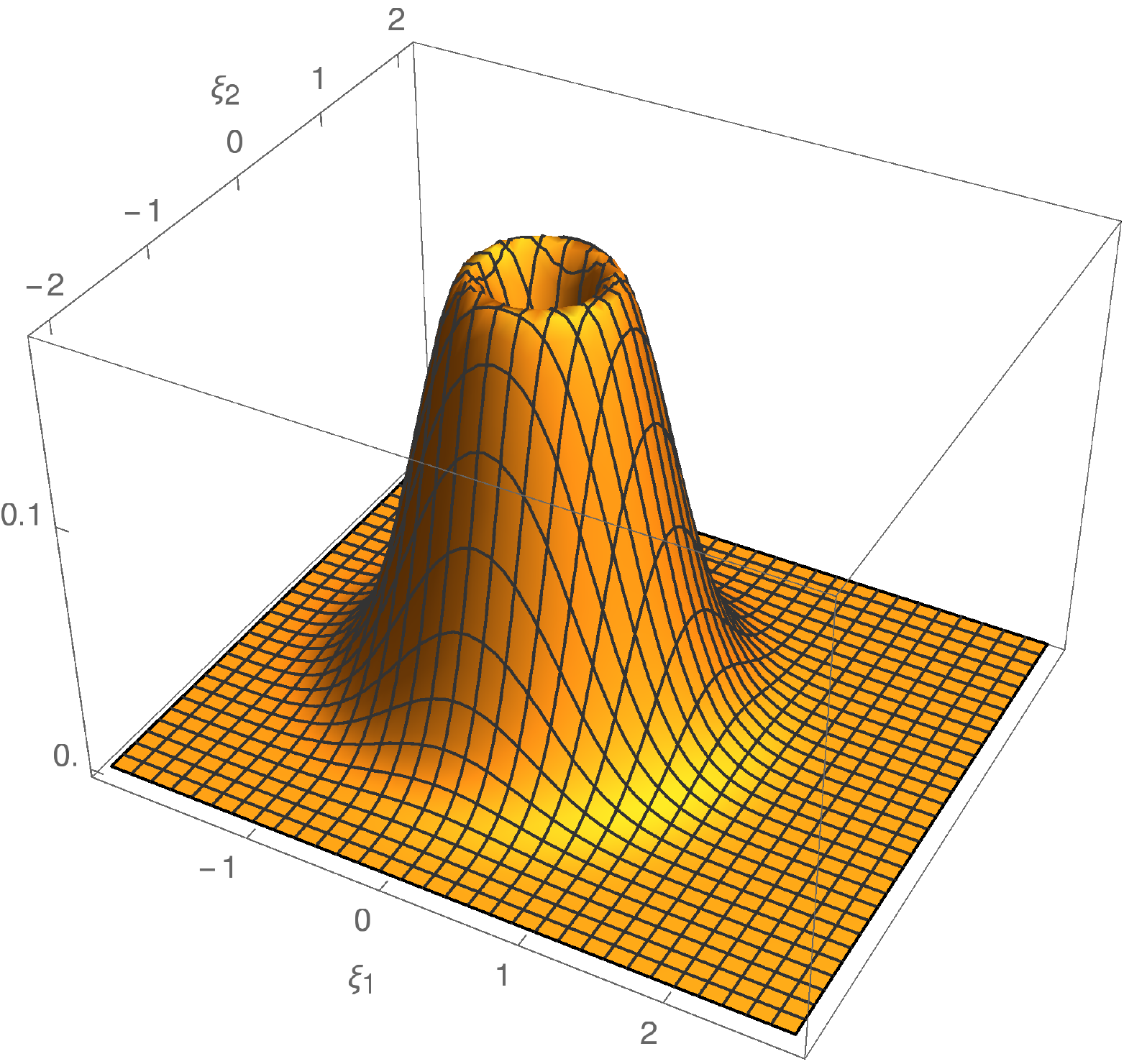} \protect\caption{\label{fig:1solenoid} The probability density $|\psi_{1}|^{2}$,
with $\psi_{1}$ being a normalized eigenfunction corresponding to
the unperturbed eigenvalue $E_{1}:=(\alpha+1/2)\hbar\omega_{c}$,
for the values of parameters $\omega_{c}=4$, $\alpha=0.4$. Here
$\pmb{\xi}:=\sqrt{\mu\omega_{c}/\hbar}\,\pmb{x}$ is dimensionless.}
\end{figure}

\clearpage{}

\begin{figure}
\includegraphics[width=0.95\textwidth]{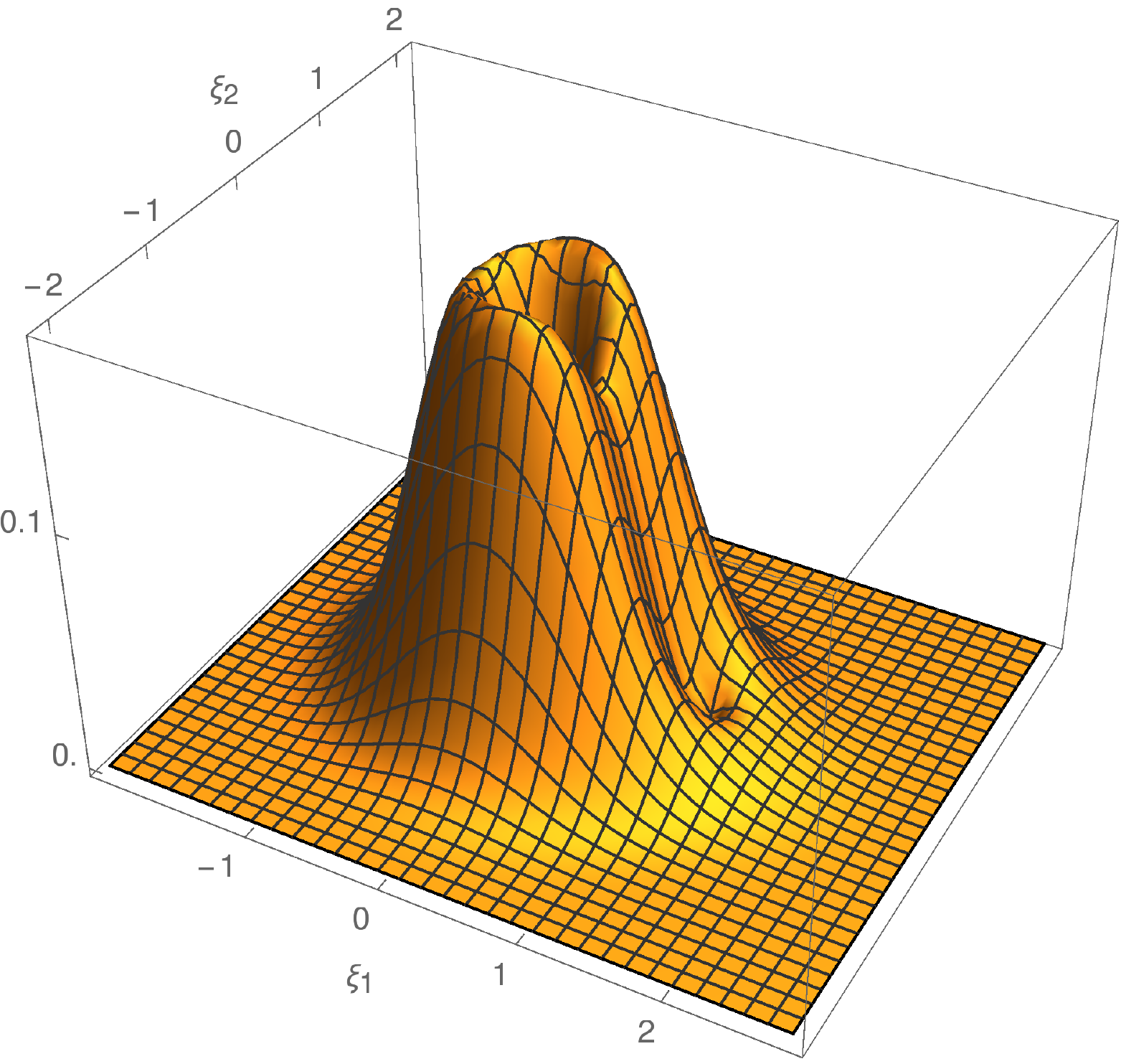}
\protect\caption{\label{fig:2solenoid} The probability density $|\psi_{2}|^{2}$,
with $\psi_{2}$ being a normalized eigenfunction corresponding to
the perturbed eigenvalue $E_{2}$ located near $E_{1}$, for the values
of parameters $\omega_{c}=4$, $D:=\mu R^{2}\omega_{c}/\hbar=3.5$,
$\alpha=0.4$, $\beta=0.7$. Again, $\pmb{\xi}:=\sqrt{\mu\omega_{c}/\hbar}\,\pmb{x}$.}
\end{figure}

\clearpage{}

\begin{figure}
\includegraphics[width=0.95\textwidth]{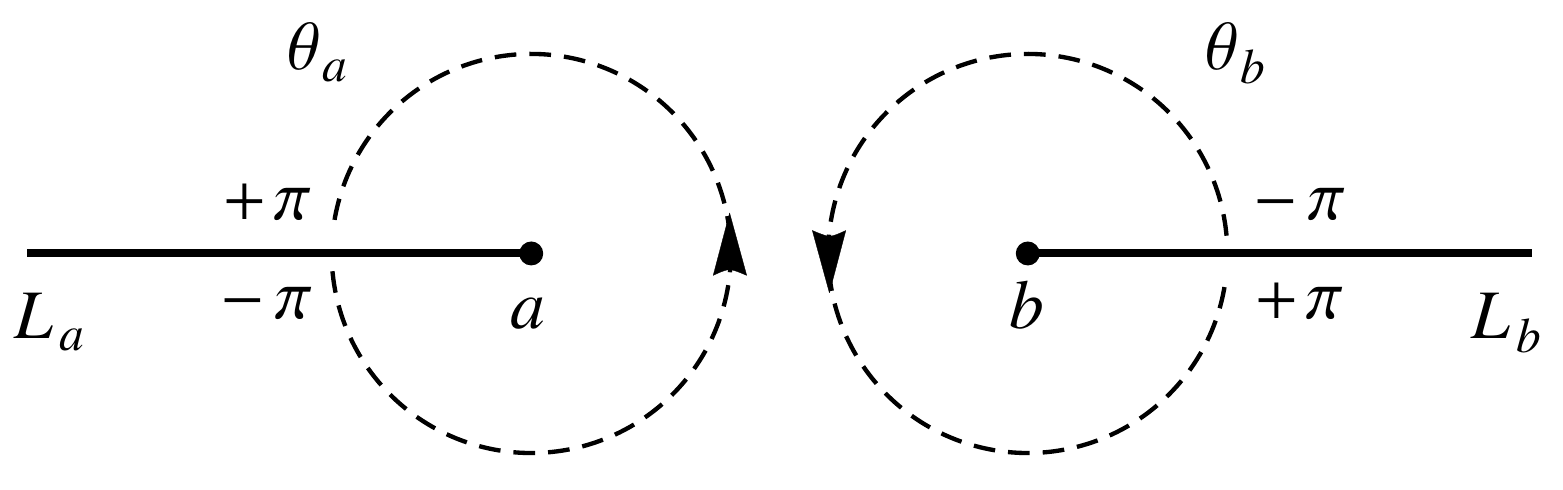} \protect\caption{\label{fig:geometry} The geometric arrangement with the AB vortices
$\pmb{a}$ and $\pmb{b}$, and the cuts of the plane $L_{a}$ and
$L_{b}$. Angle variables $\theta_{a},\theta_{b}\in(-\pi,\pi]$ are
taken with respect to the centers $\pmb{a}$ and $\pmb{b}$, respectively.}
\end{figure}

\clearpage{}
\end{document}